\documentclass[12pt]{iopart}
\usepackage{graphicx}
\usepackage{dcolumn}
\usepackage{bm}
\usepackage{iopams}
\usepackage{color}

\def\be{\begin{equation}}
\def\ee{\end{equation}}
\def\ba{\begin{array}}
\def\ea{\end{array}}
\def\bea{\begin{eqnarray}}
\def\eea{\end{eqnarray}}
\begin{document}
\title[\underline{J. Phys. G: Nucl. Part. Phys. \hspace {5.9cm} T. R. Routray et al. }]
{Exact versus Taylor-expanded energy density in the study of the neutron star crust-core transition}
\author{T. R. Routray$^{1*}$, X. Vi\~nas$^2$, D. N. Basu$^3$, S. P. Pattnaik$^1$, M. Centelles$^2$, L. Robledo$^4$ and B. Behera$^1$}
\address{$^1$School of Physics, Sambalpur University, Jyotivihar-768 019, India.}
\address{$^2$Departament de F\'{\i}sica Qu\`antica i Astrof\'{\i}sica
and Institut de Ci\`encies del Cosmos (ICCUB),
Facultat de F\'{\i}sica, Universitat de Barcelona,
Diagonal 645, E-08028 Barcelona, Spain}
\address{$^3$Variable Energy Cyclotron Center, 1/AF Bidhan Nagar, Kolkata, 700064, India}
\address{$^4$Departamento de F\'{\i}sica Te\'orica (M\'odulo15), Universidad Aut\'onoma de Madrid,
E-28049, Spain}
\ead{$^*$ trr1@rediffmail.com (corresponding author)}
\date{\today}

\begin{abstract}
The importance of the fourth and higher order terms in the Taylor series expansion of 
the energy of the isospin asymmetric nuclear matter in the study of the neutron star crust-core phase 
transition is investigated using the finite range simple effective interaction.
Analytic expressions  for the  evaluation of the second
and fourth order derivative terms in the Taylor series expansion for any general
finite range interaction of  Yukawa, exponential or Gaussian form have been obtained.
The effect of the nuclear matter incompressibility, symmetry energy and slope 
parameters on the predictions for the crust-core transition density is examined. 
The crustal moment of inertia is calculated and the prediction for the radius of the Vela pulsar is analyzed using different equations of state.
\end{abstract}



\submitto{\jpg}
\maketitle

\bigskip

\section{Introduction}
The study of the equation of state (EOS) of a neutron star (NS) is an area of current 
research interest in nuclear astrophysics. There is a lack of consensus among the 
various model calculations about the presence of deconfined quarks and hyperons 
inside the NS core \cite {lat14}. However, there is agreement in the fact that the structure of NSs 
consists of a solid crust encompassing the dense homogeneous core, which is maintained 
in a charge neutral $\beta-$stable condition and is in a liquid phase \cite {sap83}. 
The structure of the inner crust is inhomogeneous. It is made of clusters of
positive charges in a solid lattice immersed in a bath of neutrons in the
superfluid phase, along with a background of electrons such that charge neutrality is maintained. 
The complicated inhomogeneous structures often referred to as ``nuclear pasta" 
\cite {bayma,rav83,lor93,oya93,cent15} makes the 
formulation of the EOS in this region a formidable task and the value of the 
transition density between the crust and the core, $\rho_{t}$, still remains uncertain. Although the NS crust
is a small fraction of the star mass and radius, it plays an important role in various 
observed astrophysical phenomena \cite {bayma,baymb,peth95,link99,lat00,lat01,lat07,
stein05,cham08,sot12,new13,pie14,pon13,hor15}.

 The microscopic formulations of the inhomogeneous phase \cite{Dou07,Newt09,Sebi11,Schu13,Bao15}
favour a continuous nature of the transition between the crust and the 
core.
The density at which the crust-core transition takes place can be 
estimated by examining the
onset of violation of the stability condition of the homogeneous
liquid core against small-amplitude density fluctuations, which indicates
the formation of nuclear clusters. 
The three more common approaches to study the crust-core transition
are the dynamical method \cite {bayma,baymb,peth95,pet95,dou00,dou01,duc07,xu09}, the thermodynamical method
\cite{lat07,kub04,kub07,wor08,mous10,xu09,mus12} and the random phase approximation \cite {car03,hor01}. The dynamical and thermodynamical
methods are similar in the long wavelength limit giving similar 
results, but with a slightly smaller value of the transition density in the former case due to the inclusion
of density gradient and Coulomb effects \cite{xu09}. The inequalities resulting from the
stability conditions show a direct connection with the  isospin part of the EOS.
The isospin dependent part of the energy per particle 
in isospin asymmetric nuclear matter (ANM) is considered as the symmetry 
energy. The isospin dependence appears in a complicated way in the analytic formulation of the 
EOS of ANM in any of the model calculations. 
 In order to have analytical simplifications, often the Taylor series expansion of 
the energy density around zero isospin asymmetry is adopted and the energy density in
ANM is expressed as a combination of even powers of the isospin asymmetry parameter. 
In  the many-body model calculations using the Taylor series expansion of the energy density, it has been found that
higher order terms beyond $2^{nd}$ order have a
very small contribution \cite {bao08}. This finding leads to the popular parabolic approximation of
the energy per particle in ANM. Under this parabolic law, the symmetry energy
can be expressed either from the $2^{nd}$ order term in the Taylor 
series expansion of the energy density or by the empirical relation defined as the difference between the
energy per particle in pure neutron matter (PNM) and in symmetric nuclear matter (SNM). 

Although the parabolic approximation is sufficient 
for the energy calculations, the $4^{th}$ and higher order 
terms in the isospin asymmetry can have some influence when the
convergence of the Taylor series is rather slow, as for example in the calculation
of the proton fraction in $\beta-$equilibrated $n+p+e+\mu$ matter. Thus the contributions 
of these higher order terms may be significant in deciding as to whether 
the direct URCA process occurs or not in NSs \cite {ste06}.
The prediction of the crust-core transition
density $\rho_{t}$ in NSs is another important area where the role of the
higher order terms is quite significant and the results with the parabolic 
approximation can be misleading.
There are a limited number of studies \cite {xu09,mus12,cai12,sei14}, using
either non-relativistic or relativistic mean field model calculations, aimed at
explicitly demonstrating the important influence of $4^{th}$ and higher order terms
in the prediction of the critical transition density. 
A general trend that emerges from these calculations is that the inclusion of the $4^{th}$ order term reduces the value of the crust-core transition density.
This feature points out to the fact that the complete isospin contribution is 
crucial in the calculation of the transition density. 
In non-relativistic model calculations of the crust-core transition using Skyrme 
type interactions, the study 
of the influence of the $4^{th}$ and higher order terms in the Taylor series 
expansion of the  energy can be easily performed due to the simple form 
of the analytic expressions. However, for 
finite range interactions the complexity increases. 
In  Ref.\cite {xu09}, the influence of higher order terms of the 
Taylor series expansion on the crust-core transition and the properties of a
NS has been studied using the MDI energy density \cite {das03} that can
result from a finite range interaction having a Yukawa form. In the present work,
we have obtained general expressions that can be used to obtain the 
$2^{nd}$ and $4^{th}$ order contributions of the Taylor series expansion 
for any conventional form of  finite range interactions. 
The values of
$\rho_{t}$ and the pressure at the transition density, $P(\rho_{t})$, are the
quantities of crucial importance in the calculation of the crustal fraction of the
moment of inertia $\Delta{I}/{I}$, where $I$ is the moment of inertia of the NS 
\cite {link99,xu09}. The quantity $\Delta{I}/{I}$ can be estimated from the observations of pulsar glitches. Glitches are the intermittent sudden jumps in the rotational
frequencies of pulsars, which are believed to arise from the interaction between
the solid crust and the liquid core. 

In section \ref{Sec:formalism}, we outline the stability conditions of the thermodynamical 
method in terms of the nuclear EOS. Analytic expressions for the $2^{nd}$ and $4^{th}$ 
order terms in the Taylor series expansion of the energy density in ANM have been obtained. 
The explicit expressions for the energy per particle in ANM, neutron (proton) chemical potential 
and other quantities used in the study 
have been obtained for the Yukawa form of the finite range simple effective interaction (SEI)
and are given in the Appendix. 
 Alternatively, and for practical purposes, these expressions 
can be evaluated numerically (including thermodynamic derivatives). 
The procedure adopted to determine the parameters of SEI
necessary for the study of ANM is outlined. The approximate expression of the crustal momentum 
of inertia obtained in Ref. \cite {link99,xu09} is outlined in this 
section. Section \ref{Sec:res} contains results and discussions on the transition 
density obtained with the 
Yukawa form of the SEI for the $2^{nd}$ and $4^{th}$
order Taylor 
series, as well as for the exact analytic case. We discuss the importance of using exact expressions of the energy density
instead of their Taylor expansions to compute the crust-core transition density.  
The influence of nuclear matter 
saturation properties such as incompressibility, symmetry energy and slope 
parameter on the transition density are also discussed. The crustal moment of 
inertia as well as the  mass-radius relation for the Vela 
pulsar are calculated for the EOSs under consideration. The influence of the functional form of the finite range part
of the interaction on the transition density is also examined by considering a Gaussian form factor. 
Finally, Section \ref{Sec:con} contains the final remarks and conclusions.

\section{Formalism}
\label{Sec:formalism}

The study of the crust-core phase transition in neutron stars in the
thermodynamical model amounts to examine the stability of the $ \beta-$equilibrated charge neutral 
dense matter in the liquid phase with respect to the stability conditions \cite {kub04,kub07}, 
\begin{eqnarray}
-\left(\frac{\partial P}{\partial v}\right)_{\mu} >{0}
\label{eq1}
\end{eqnarray}
\begin{eqnarray}
-\left( \frac{\partial \mu }{\partial q }\right)_{v}>0.
\label{eq2}
\end{eqnarray}
In these equations,
$P$ is the total pressure, $v= V/B$ is the volume per baryon number, $q$ is the charge fraction
given by  $q=Y_{p}-Y_{e}$, where $Y_{i}=\rho_{i}/\rho$, $i=p,e$, are the proton and electron
fractions, with $\rho=\rho_{p}+\rho_{n}$ being the
total nucleonic density and $\rho_{n}$, $\rho_{p}$ and $\rho_{e}$ being the neutron, proton and
 electron densities, respectively. The $\beta$-equilibrium chemical potential is
$\mu=\mu_{n}-\mu_{p}=\mu_{e}$, $\mu_{i}$, where $i=n,p,e$ denotes the respective chemical potentials.
The stability conditions in equations (1) and (2) refer to the mechanical and chemical stabilities
of the system, respectively. Here, the total pressure $P$ of the ($n$,$p$,$e$) system comprises
the baryonic contribution $P^{N}$ and the leptonic contribution $P^{l}$. Under the consideration
of the relativistic Fermi gas model, the pressure of the electrons is $P^{l}=\frac{\mu^{4}_{e}}{12 \pi^{2}(c\hbar)^{3}}$.
The electronic contribution to the pressure becomes a function of $\mu$ only when the system is in
$\beta$-equilibrium and does not contribute to the stability condition in equation (1), which now reads
\begin{eqnarray}
-\left(\frac{\partial P^{N}}{\partial v}\right)_{\mu} >{0} .
\label{eq3}
\end{eqnarray}
From the relation  $q=Y_{p}-Y_{e}=Y_{p}-\rho_{e}/\rho$ , where $\rho_{e}=\frac{\mu^{3}_{e}}{3\pi^{2}c^{3}\hbar^{3}}$,
it follows that
\begin{eqnarray}
-\left(\frac{\partial q}{\partial \mu}\right)_{v} = - \left(\frac{\partial Y_{P}}{\partial\mu}\right)_{v}+
 \frac{1}{\rho}\left(\frac{\partial \rho_{e}}{\partial \mu}\right)_{v} .
\label{eq4}
\end{eqnarray}
Now using the relations $v=\frac{1}{\rho}$ and $\mu=-\frac{\partial e(\rho,Y_{p})}{\partial Y_{p}}$,
where $e(\rho,Y_{P})$ is the energy per nucleon in the $\beta$-equilibrated nuclear matter ($\mu=\mu_{e}$),
the conditions in equations (3) and (4) identically reduce to \cite{mous10,xu09,att14}
\begin{eqnarray}
\rho^{2}\left(\frac{\partial P^{N}}{\partial \rho}\right)_{\mu}\nonumber\\
= 2\rho^{3} \left(\frac{\partial\ e(\rho,Y_{p})} 
{\partial\rho} \right) + \rho^{4} \left (\frac {\partial^{2} e(\rho,Y_{p})}
{\partial\rho^{2}}\right)-\rho^{4}\frac{\left( \frac{\partial^{2}e(\rho,Y_{p})}{\partial\rho \partial Y_{p}}\right)^{2}}
 {\left ({\frac{\partial^{2}e(\rho,Y_{p})}{\partial Y_{p}^{2}}}\right)}>0 ,
\label{eq5}
\end{eqnarray}
\begin{eqnarray}
-\left( \frac{\partial q }{\partial \mu }\right)_{v} = \left(\frac{\partial^{2}e(\rho,Y_{p})}{\partial Y^{2}_{P}} \right)^{-1}
+\frac{\mu^{2}}{\pi^{2}(c\hbar)^{3} \rho}>0 .
\label{eq6}
\end{eqnarray}
In the second condition one may recognize the last term in the right hand side (rhs) as the screening length.
The first term in the rhs can be shown, under the parabolic approximation, to be proportional to nuclear symmetry energy
$E_{s}(\rho)$.
Although the density behaviour of
$E_{s}(\rho)$ is still uncertain, the heavy-ion collision studies favour a monotonically increasing behaviour of
$E_{s}(\rho)$ and therefore the condition in equation (6) is usually satisfied. The first two terms in the rhs of 
equation (5) refer to the pressure and incompressibility of nucleonic matter and are positive for fundamental reason, whereas
the third term which comes from the presence of leptons contributes negatively. Thus the stability condition in equation (5)
reduces to the form
\begin{eqnarray}
V_{thermal}= 2\rho \left(\frac{\partial\ e(\rho,Y_{p})} {\partial\rho} \right) + \rho^{2} \left (\frac {\partial^{2} 
e(\rho,Y_{p})} 
{\frac{\partial^{2}e(\rho,Y_{p})}{\partial Y_{p}^{2}}}\right)
-\frac{\left(\rho \frac{\partial^{2}e(\rho,Y_{p})}{\partial\rho \partial Y_{p}}\right)^{2}}
{\left ({\frac{\partial^{2}e(\rho,Y_{p})}{\partial Y_{p}^{2}}}\right)}>0 ,
\label{eq7}
\end{eqnarray}
where $V_{thermal}= \left(\frac{\partial P^{N}}{\partial \rho}\right)_{\mu}$ and $e(\rho,Y_{p} )=\frac{H(\rho,Y_{p})}{\rho}$
is the energy per
particle in ANM, $H(\rho,Y_{p})$ being the energy density of ANM.
From the relation in equation (\ref{eq7}) the importance of the EOS of ANM in
determining the transition density for the crust-core transition in a NS is evident.
 Once the equilibrium proton fraction $Y_{p}$ is ascertained for a given nucleonic density $\rho$
from the solution of the charge neutral beta stability conditions, the nucleonic energy per particle
$e(\rho,Y_{p} )$ can be calculated and $V_{thermal}$ evaluated using equation (\ref{eq7}) 
either numerically or by analytical procedure wherever it is possible. Numerical evaluation 
of equation (\ref{eq7}), however, does not provide further insight into the isospin dependence of 
the EOS of ANM which is an area that is less understood and of current research interest.

The analytical calculation of $V_{thermal}$ from equation (\ref {eq7}) can be conveniently
made provided the
$Y_p$-dependence in the expression for the energy density in ANM is separated out. The isospin 
dependence in the energy density of ANM resulting from a finite range interaction is 
complicated and in most cases it is not practically possible to express the energy density by segregating the $Y_p$-dependent
parts. Due to this, the Taylor series expansion of the energy density of ANM is usually adopted in the studies 
involving the isospin dependence aspects. For calculating the exact
analytic expression of $V_{thermal}$, equation (\ref {eq7}) can be equivalently expressed
in terms of derivatives of the chemical potentials $\mu_{q}$ $(q=n,p)$ with respect to the neutron and proton densities,
$\rho_n$ and $\rho_p$, as given by
\begin{eqnarray}
V_{thermal}&=\frac{\rho}{4}
\Bigg[ 
  \left( 
  \frac{\partial \mu_{n}}{\partial \rho_{n}}
+2\frac{\partial \mu_{n(p)}}{\partial \rho_{p(n)}}
+ \frac{\partial \mu_{p}}{\partial \rho_{p}}
  \right)
+2(1-2Y_{p}) 
 \left(
 \frac{\partial \mu_{n}}{\partial \rho_{n}}
-\frac{\partial \mu_{p}}{\partial \rho_{p}}
 \right) \nonumber\\
&+(1-2Y_{p})^{2}
  \left(
  \frac{\partial \mu_{n}}{\partial \rho_{n}}
-2\frac{\partial \mu_{n(p)}}{\partial \rho_{p(n)}}
+ \frac{\partial \mu_{p}}{\partial \rho_{p}} 
  \right)\nonumber\\
&-\frac{\left\{  (\frac{\partial \mu_{n}}{\partial \rho_{n}} -\frac{\partial \mu_{p}}{\partial \rho_{p}})+(1-2Y_{p})(\frac{\partial \mu_{n}}{\partial \rho_{n}}-2\frac{\partial \mu_{n(p)}}{\partial \rho_{p(n)}} + \frac{\partial \mu_{p}}{\partial \rho_{p}})          \right\}^{2}} {\frac{\partial \mu_{n}}{\partial \rho_{n}}-2\frac{\partial \mu_{n(p)}}{\partial \rho_{p(n)}}
+ \frac{\partial \mu_{p}}{\partial \rho_{p}}}
\Bigg]>0 ,
\label{eq8}
 \end{eqnarray}
where $\mu_{i}=\frac {\partial H(\rho,Y_p)}{\partial \rho_{i}}$ for $i=n, p$.
Further, the equality $\frac {\partial \mu_{n}}{\partial \rho_{p}}=\frac {\partial \mu_{p}}{\partial \rho_{n}}$
is used. 
The exact isospin dependence of the EOS of ANM is poorly known and a popular
way of studying it is by making a Taylor series expansion of the energy density
in ANM in even powers of the isospin asymmetry parameter, $\beta=\frac{\rho_{n}-\rho_{p}}{\rho_{n}+\rho_{p}}=(1-2Y_{p})$, 
\begin{eqnarray}
H(\rho,Y_{p})=H(\rho) +(1-2Y_{p})^{2} H_{sym,2}(\rho)+(1-2Y_{p})^{4} H_{sym,4}(\rho)+...
\label{eq9}
 \end{eqnarray}
with,
\begin{eqnarray}
 H_{sym,2}(\rho)=\left(\frac{1}{8} \frac{\partial^{2}H(\rho,Y_{p})}{\partial Y_{p} ^{2}} \right)_{Y_{p}=1/2}
 \label{eq10}
 \end{eqnarray}
\begin{eqnarray}
H_{sym,4}(\rho)=\left(\frac{1}{384} \frac{\partial^{4}H(\rho,Y_{p})}{\partial Y_{p} ^{4}} \right)_{Y_{p}=1/2}
 \label{eq11}
 \end{eqnarray}
where $H(\rho)=\rho\, e(\rho)$ is the energy density in SNM.
The analytic expressions of these $2^{nd}$ and $4^{th}$ order terms for 
any general effective interaction, $v(\bf r)$, are given as,
\begin{eqnarray}
&H_{sym,2}(\rho,Y_{p})=\frac{\hbar^{2}k^{2}_{f} \rho}{6m}+\frac{\rho^{2}}{4} \int \left( v^{l}_{d}(r)-v^{ul}_{d}(r) \right) d^{3}r\nonumber\\
 &+\frac{\rho^{2}}{4} \left[ \int \left( v^{l}_{ex}(r)-v^{ul}_{ex}(r) \right)j^{2}_{0}(k_{f}r) d^{3}r - 
 \int \left( v^{l}_{ex}(r)+v^{ul}_{ex}(r) \right)j^{2}_{1}(k_{f}r) d^{3}r    \right]
\quad
 \label{eq12}
 \end{eqnarray}
and
\begin{eqnarray}
 H_{sym,4}(\rho,Y_{p})&=\frac{\hbar^{2}k^{2}_{f} \rho}{162m}\nonumber\\
 &+\frac{\rho^{2} k^{2}_{f}}{108} \int \left( 4 \frac{ j_{1}(k_{f}r)}{(k_{f}r)} - j_{0}(k_{f}r)  \right) j_{0}(k_{f}r) (v^{l}_{ex}(r)-v^{ul}_{ex}(r) ) r^{2} d^{3}r\nonumber\\
&+\frac{\rho^{2} k^{2}_{f}}{108} \int (v^{l}_{ex}(r)+v^{ul}_{ex}(r) ) j^{2}_{1}(k_{f}r) r^{2} d^{3}r\nonumber\\
&+\frac{\rho^{2} k_{f}}{54}\int (v^{l}_{ex}(r)+v^{ul}_{ex}(r) )j_{0}(k_{f}r) j_{1}(k_{f}r) r d^{3}r\nonumber\\
&-\frac{7 \rho^{2} }{108}\int (v^{l}_{ex}(r)+v^{ul}_{ex}(r) )j^{2}_{1}(k_{f}r)  d^{3}r
 \label{eq13}
 \end{eqnarray}
respectively. In these expressions $k_{f}=(3\pi^{2}\rho/2)^{1/3}$ is the Fermi momentum in SNM, $v_{d}^{l}(r)$, $v_{d}^{ul}(r)$ and $v_{ex}^{l}(r)$, $v_{ex}^{ul}(r)$
are the direct (d) and exchange (ex) parts of the interaction acting between 
like (l) and unlike (ul) pairs of nucleons, $m$ is the nucleon mass and $j_{0}$ and $j_{1}$ are the spherical Bessel
functions of zeroth and first order.
In all the model calculations it is found that the contributions  to the energy from the
higher order terms, beyond the $2^{nd}$ order, is very small and therefore
the series is often truncated at the quadratic term resulting into the parabolic
expression of the energy per particle in ANM, 
\begin{eqnarray}
 e(\rho,Y_{p})=e(\rho) + (1-2Y_{p})^{2} {E_{s}(\rho)},
 \label{eq14}
 \end{eqnarray}
where the symmetry energy $E_{s}(\rho)$ is obtained from the second order
derivative term as $H_{sym,2}(\rho)/\rho $. 
Another empirical expression of the symmetry energy $E_{s}(\rho)$, often used, 
is by approximating it as the difference between the energy per particle 
in PNM, $e^{N}(\rho)$, and in SNM, $e(\rho)$, 
 \begin{eqnarray}
E_{s}(\rho)= e^{N}(\rho) -e(\rho) .
 \label{eq15}
 \end{eqnarray}
The justification of this approximation is that at the two extreme values of the
isospin asymmetry, $Y_{p}$=0 and $Y_{p}$=1/2, the energy per particle expression 
$e(\rho,Y_{p})=e(\rho)+(1-2Y_{p})^{2} {E_{s}(\rho)}$ of ANM reduces to the 
respective expressions of PNM and SNM, respectively.
 
For the $2^{nd}$ order approximation of the energy density in the Taylor series 
expansion, the stability condition $V_{thermal}$ in equation (\ref {eq7}) becomes
\cite {xu09,cai12,att14},
\begin{eqnarray}
 V_{thermal}&=\rho^{2} \frac{ \partial^{2} e(\rho)}{\partial \rho^{2}} + 2 \rho \frac{ \partial e (\rho)}{\partial \rho }
  +\beta^{2} \left[\rho^{2}   \frac{ \partial^{2} E_{sym,2}(\rho)}{\partial \rho^{2}} \right. \nonumber\\
 & \left. +2 \rho \frac{ \partial E_{sym,2}(\rho)}{\partial \rho }
 -\frac{2 \rho^2}{E_{sym,2}(\rho)}\Big(\frac{ \partial E_{sym,2}(\rho)}{\partial \rho }\Big)^{2} \right]>0,
 \label{eq16}
 \end{eqnarray}
where $E_{sym,2}=\frac {H_{sym,2}}{\rho}$. Similarly, for the $4^{th}$ order approximation, where the Taylor series
expansion in equation (\ref {eq9}) is truncated at the fourth order 
term, the stability condition becomes \cite {cai12,sei14}, 
 \begin{eqnarray}
 V_{thermal}&=\rho^{2} \left( \frac{ \partial^{2} e(\rho)}{\partial \rho^{2}}\right) + 2 \rho\left( \frac{ \partial e(\rho)}{\partial \rho }\right)
\nonumber\\
 &+ \beta^{2}\left[\rho^{2}\left(  \frac{ \partial^{2} E_{sym,2}(\rho)}{\partial \rho^{2}}\right) + 2 \rho\left( \frac{ \partial
 E_{sym,2}(\rho)}{\partial \rho }\right)\right]\nonumber\\
 &+\beta^{4}\left[\rho^{2}\left(  \frac{ \partial^{2} E_{sym,4}(\rho)}{\partial \rho^{2}}\right) + 2 \rho\left( \frac{ \partial E_{sym,4}(\rho)}{\partial \rho }\right)\right]\nonumber\\
&-\frac{2\beta^{2}\rho^{2}}{E_{sym,2}(\rho) +6 \beta^{2} E_{sym,4}(\rho)} \left[  \frac{ \partial E_{sym,2}(\rho)}{\partial \rho }+ 2 \beta^{2} \frac{\partial E_{sym,4}(\rho)}{\partial \rho}  \right]^{2}>0
, \;\;\;
\label{eq17}
  \end{eqnarray}
where $E_{sym,4}=\frac {H_{sym,4}}{\rho}$.
The difference between the neutron and proton chemical potentials in the $\beta-$ stability condition,
\begin{eqnarray}
\mu_{n}-\mu_{p}=\mu_{e}=\mu_{\mu},
\label{eq18}
  \end{eqnarray}
for the $2^{nd}$ and $4^{th}$ order Taylor series approximations of the energy 
density becomes, 
 \begin{eqnarray}
\mu_{n}-\mu_{p}=4  (1-2Y_{p}) E_{sym,2} (\rho)
\label{eq19}
  \end{eqnarray}
and
 \begin{eqnarray}
\mu_{n}-\mu_{p}=4(1-2Y_{p}) E_{sym,2} (\rho)+8(1-2Y_{p})^{3}E_{sym,4} (\rho) ,
\label{eq20}
  \end{eqnarray}
respectively.

 The stability condition of $V_{thermal}$ is a signature of the crust-core transition in the $\beta-$equilibrated neutron star matter. As we summarize in the next subsection, the transition density and the pressure at the transition density play a critical role in the prediction of the crustal fraction of the moment of inertia of neutron stars, used in the possible explanation of the observed glitches in pulsars.

\subsection{\it Crustal fraction of the moment of inertia of neutron stars}

Based on the hypothesis that the mechanism for glitches observed in the
magnetized rotating neutron stars, is due to the pinning of the vortexes in the superfluid neutrons inside the dense liquid core with the superfluid
neutrons of the inner crust in the crust-core transition region \cite {eps92,lin96,rud98,sed99},
the crustal fraction of moment of inertia, $\Delta {I}/{I}$, can be
calculated from the observed glitches. Glitches are the intermittent disruption in the
extremely regular pulses emitted from the magnetized rotating neutron stars. 
An approximate 
expression for $\Delta {I}/{I}$ has been obtained by Xu {\it et. al.} \cite {xu09} using the work of Link {\it et. al.} \cite {link99}. It contains the mass $M$ and  radius $R$ of the
NS, and the dependence on the EOS through the pressure and the
transition density at the crust-core transition, $P(\rho_{t})$ and $\rho_{t}$, 
respectively, as given by
\begin{eqnarray} 
\frac {\Delta{I}}{I} &\approx \frac{28\pi P({\rho_t})R^{3}}{3Mc^{2}} \left( \frac{1-1.67\xi-0.6\xi^{2}}{\xi} \right)\nonumber\\
&\times \left(1+\frac{2P({\rho_t})}{\rho_{t}mc^{2}} \frac{(1+7\xi)(1-2\xi)}{\xi^{2}} \right)^{-1} ,
 \label{eq21}
  \end{eqnarray}
where $\xi=\frac {GM}{Rc^2}$, $G$ is the gravitational constant and
$c$ is the velocity of light. The NS mass and radius for a given EOS are calculated by solving the 
Tolman-Oppenheimer-Volkov (TOV) equations. 
The total moment of inertia, $I$, of the neutron star rotating slowly with a 
uniform angular velocity $\Omega$ is obtained from $I=J/{\Omega}$, 
where the total angular momentum $J$ is calculated from \cite {arn77},
\begin{eqnarray}
J=\frac{c^{2}}{6G} R^{4} \frac{d \bar {\omega}}{dr}\Big|_{r=R} .
\label{eq22}
 \end{eqnarray}
The angular velocity of a point in the star, $\bar {\omega}$, is obtained
from the solution of the relevant equation in Ref. \cite {arn77} subject to the 
boundary conditions that $\bar {\omega}$ is regular at $r \rightarrow 0$ and 
$\bar {\omega}\rightarrow \Omega$ as $r \rightarrow \infty$.

\subsection {Finite range simple effective interaction}

The present study is made using the finite range simple effective interaction (SEI)  described in \cite {trr98,trr02}:
\begin{eqnarray}
v_{eff}(\mathbf{r})&=&t_0 (1+x_0P_{\sigma})\delta(\mathbf{r}) \nonumber \\
&&+\frac{t_3}{6}(1+x_3 P_{\sigma})\left(\frac{\rho({\bf R})}
{1+b\rho({\bf R})}\right)^{\gamma} \delta(\mathbf{r}) \nonumber \\
&& + \left(W+BP_{\sigma}-HP_{\tau}-MP_{\sigma}P_{\tau}\right)f(\mathbf{r}),
\label{eq23}
\end{eqnarray}
 where $\mathbf{r}=\vec{r}_1-\vec{r}_2$, $\mathbf{R}=(\vec{r}_1+\vec{r}_2)/2$ and $f(\mathbf{r})$ is the functional form of the finite range  interaction, which may
be of Yukawa, Gaussian or exponential form, and contains a single range
parameter $\alpha$. The SEI contains altogether eleven parameters, namely,
$t_0$, $x_0$, $t_3$, $x_3$, $b$, $\gamma$, $\alpha$, $W$, $B$, $H$ and $M$.
However, for the study of isospin asymmetric nuclear matter only nine parameters are required, namely, $b$, $\gamma$, $\alpha$, 
$\varepsilon_{0}^{l}$, $\varepsilon_{0}^{ul}$, 
$\varepsilon_{ex}^{l}$,
$\varepsilon_{ex}^{ul}$, $\varepsilon_{\gamma}^{l}$, 
$\varepsilon_{\gamma}^{ul}$. The connection of the new parameters to the
interaction parameters is given in Ref.\cite {trr13}. In this work we shall
be calculating the results for the Yukawa form of $f(r)$.
The determination of the 
nine parameters of the ANM is discussed in detail in earlier studies
\cite {trr13,trr07,trr15}. Here we outline the procedure for the sake of convenience of the reader.

The \textsl{symmetric} nuclear matter requires only the  following combinations
of the strength parameters in the like and unlike channels:
\begin{eqnarray}
\left(\frac{\varepsilon_{0}^{l}+\varepsilon_{0}^{ul}}{2}\right)=\varepsilon_0,
\left(\frac{\varepsilon_{\gamma}^{l}+\varepsilon_{\gamma}^{ul}}{2}\right)=\varepsilon_{\gamma},
\left(\frac{\varepsilon_{ex}^{l}+\varepsilon_{ex}^{ul}}{2}\right)=\varepsilon_{ex},
\label{eq24}
\end{eqnarray}
together with the $\gamma$, $b$ and $\alpha$ parameters, altogether six 
parameters. 
For a given value of the exponent
$\gamma$, and assuming 
the standard values for the nucleon mass $mc^{2}$, saturation density $\rho_{0}$ and 
binding energy per particle in SNM at saturation $e(\rho_{0})$, 
the  remaining five parameters $\varepsilon_{0}$, $\varepsilon_{\gamma}$,
$\varepsilon_{ex}$, $b$ and $\alpha$ 
of symmetric nuclear matter are determined in the following way. 
The range $\alpha$ and the exchange strength $\varepsilon_{ex}$ are determined 
simultaneously by adopting an optimization procedure using the condition 
that the nuclear mean field in symmetric nuclear matter at saturation density 
vanishes for  a nucleon kinetic
energy of 300 MeV, a result extracted from optical model analysis
of nucleon-nucleus data \cite{bers88,gale90,cser92}. 
This requires only the values of $mc^{2}$, $\rho_{0}$ and $e(\rho_{0})$ 
as discussed in Refs \cite{trr98,trr02} and is independent of the other parameters
of symmetric nuclear matter including $\gamma$.
The parameter $b$ is fixed independently for avoiding the
supraluminous behaviour in SNM \cite{trr97}.
The two remaining parameters,
namely $\varepsilon_{\gamma}$ and $\varepsilon_{0}$, are obtained from the saturation
conditions. The stiffness parameter $\gamma$ is kept as a free parameter 
and its allowed values are chosen subject to the condition that the
pressure-density relation in symmetric matter lies within the region 
extracted from the analysis of flow data in heavy-ion collision experiments 
at intermediate and high energies \cite{Danielz02}. It is found that the 
maximum value that fulfills this condition is $\gamma$=1, which corresponds 
to a nuclear matter incompressibility $K(\rho_0)$=283 MeV. Therefore, we can 
study the nuclear matter properties by assuming different values of $\gamma$ 
up to the limit $\gamma$=1.

Now,
to describe \textsl{asymmetric} nuclear matter we need to know how the strength 
parameters
$\varepsilon_{ex}$, $\varepsilon_{\gamma}$ and $\varepsilon_{0}$ 
split into
the like and unlike components. The splitting of $\varepsilon_{ex}$ into
$\varepsilon_{ex}^{l}$ and $\varepsilon_{ex}^{ul}$ is decided from the 
condition that the entropy density in pure neutron matter should not exceed 
that of the symmetric nuclear matter. This prescribes the critical value for 
the splitting of the exchange strength parameter to be $\varepsilon_{ex}^{l}=2\varepsilon_{ex}/3$ \cite{trr11}.
The splitting of the remaining two strength parameters
$\varepsilon_{\gamma}$ and $\varepsilon_{0}$,
is obtained from the values of the symmetry energy $E_s(\rho_0)$ and
its derivative $E_s^{'}(\rho_0)$ = $\rho_0 \frac{dE_s(\rho_0)}{d\rho_0}$ 
at saturation density $\rho_0$. Notice that the usual slope parameter
of the symmetry energy is defined as $L=3E_s^{'}(\rho_0)= 3\rho_0 \frac{dE_s(\rho_0)}{d\rho_0}$.
By assuming a value for 
$E_s(\rho_0)$ within its accepted range \cite{dutra12,dan14}, 
we determine $E_s^{'}(\rho_0)$ from the condition that the difference 
between the energy densities of the nucleonic part in charge neutral 
beta-stable $n+p+e+\mu$ matter, referred to as neutron star matter (NSM), 
and in symmetric matter at the same density 
be maximal \cite{trr07}. The value of $E_s^{'}(\rho_0)$ thus obtained for a given $E_s(\rho_0)$ 
predicts a density dependence of the symmetry energy which is neither very stiff
nor soft and does not predict the direct URCA process in the calculated NSs. It may noted here that the splitting of the exchange strength parameter $\varepsilon_{ex}$ into the like and unlike channels in ANM is solely
determined, as discussed in Ref \cite{trr11}, from the thermal evolution of 
nuclear matter properties in SNM and PNM, and is independent of the splitting
of the parameters $\varepsilon_{0}$ and $\varepsilon_{\gamma}$. 
Thus, the parameters associated with the finite range exchange part of the mean field 
and EOS, namely $\varepsilon_{ex}$, $\alpha$, $\varepsilon^{l}_{ex}$ 
and $\varepsilon^{ul}_{ex}$ are independent of the way in which the remaining
parameters of the interaction, including $\gamma$, are determined and also of the choice of $E_{s}(\rho_{0})$.

With the parameters determined in this way, the SEI was able to reproduce 
the trends of
the EOS and the momentum dependence of the mean field properties 
with similar quality as predicted by microscopic calculations
\cite{trr11,trr09}. 
  The two sets of parameters of the EOSs of SEI corresponding to $\gamma$=1/2 
and $\gamma$=1, which cover a range of nuclear matter incompressibility between 
240 MeV and 280 MeV, are given in Table 1,
along with their nuclear matter saturation properties. The relevant analytic expressions of the various quantities required in the study for the Yukawa form of SEI are given in the Appendix.
 As it can seen from Table~1, the SEI values of the saturation density and binding energy per nucleon in SNM lie within the empirical ranges 0.17$\pm$0.03 fm$^{-3}$ and 16$\pm$0.2 MeV \cite{trr13}. The value of the symmetry energy $E_s(\rho_0)$ also agrees well with the range 29-33 MeV provided by recent analyses \cite{tsang12,lat13}. The slope parameter of the symmetry energy $L(\rho_0)$ is also compatible with the commonly accepted range between 40 and 70 MeV \cite{vinas14}. The analysis
of the excitation energy of the giant monopole resonance provides a range of allowed values
of the incompressibility $K$ between 200 and 260 MeV \cite{dutra12}. The SEI used in this work
have $K$ values around 240 MeV ($\gamma$=1/2) and 280 MeV ($\gamma$=1). Although this second value is slightly high, we use it to simulate a stiffer EOS. However, we can see in the middle panel of Figure 1 that the two considered EOS computed with the SEI used in this work lie within the boundaries of allowed values extracted from the analysis of the flow data in heavy-ion collisions \cite{Danielz02} and kaon production data \cite{lynch09}.
%
%
 In the left panel of Figure 1 we compare the energy per particle in SNM and PNM, $e(\rho)$ and $e^N(\rho)$,
respectively, as a function of the density computed using the SEI corresponding to $\gamma$=1/2 with the microscopic 
Dirac-Brueckner-Hartree-Fock calculation using the Bonn B potential \cite{sam10} and with the variational calculation 
with the $A18+\delta v+UIX^*$ realistic interaction \cite {akm98}. It can be seen that, for both SNM and PNM,
there is a good agreement between the SEI predictions and the microscopic results up to about $\rho=0.3 fm^{-3}$.
Above this density, the SNM and PNM EOSs obtained with SEI follow rather well the trend of the variational calculation 
and differ more from the DBHF results, which grows faster in SNM and PNM compared with the corresponding SEI values. 
The symmetry energy, which is approximated as the difference of $e^N(\rho)$ and $e(\rho)$, shall have similar 
matching in their density dependence. The results for the symmetry energy of the SEI EOS shall be close to the 
prediction of the realistic interaction $A18+\delta v+UIX^*$ upto density $\rho=0.75 fm^{-3}$ as can be seen from the 
the respective results of $e^N(\rho)$ and $e(\rho)$ in the {\it left} panel of figure 1. 
Another important aspect is the momentum dependence of the neutron and proton mean fields in ANM. 
The neutron (proton) effective mass, $(m^{*}/m)_{n(p)}$, as function of momentum $k$, density $\rho$ and asymmetry $\beta$, is defined as 
\begin{eqnarray}
\bigg[\frac{m^*}{m}(k,\rho,Y_{p})\bigg]_{n(p)}=\bigg[1+\frac{m}{\hbar^{2}k} 
\frac{\partial u^{n(p)}(k,\rho,Y_{p})}{\partial k}\bigg]^{-1},
\label{eq25}
\end{eqnarray}
where, $u^{n(p)}$ is the neutron (proton) mean field. This quantity can be taken as a measure of the momentum dependence of the neutron (proton) single particle potentials. 
The neutron-proton effective mass splitting at saturation density and momentum equal to the Fermi momentum at saturation 
density for the SEI having $\gamma$=1/2 is displayed as a function of the asymmetry $\beta$ in the rightmost
panel of Figure 1 compared with the microscopic DHFB \cite{sam10} and BHF \cite{zuo05} predictions. It can be seen that
the SEI results compare very well with the DBHF results over the whole range of asymmetry, while they differ more from
the BHF prediction \cite{zuo05}. This fact shows that, at least in normal nuclear matter, the momentum dependence of the 
SEI mean fields are similar to those of the {\it ab initio} DBHF formulation.  

\begin{table}
\caption{Values of the nine parameters of ANM for the two sets of SEI corresponding to $\gamma$=1/2 and $\gamma$=1 together with their nuclear matter saturation properties (see text for details).}
\renewcommand{\tabcolsep}{0.05cm}
\renewcommand{\arraystretch}{1.2}
\begin{tabular}{|c|c|c|c|c|c|c|c|c|c|c|c|}\hline
\hline
$\gamma$ & $b$    & $\alpha$ & $\varepsilon_{ex}$ & $\varepsilon_{ex}^{l}$ & $\varepsilon_{0}$ &
$\varepsilon_{0}^{l}$ & $\varepsilon_{\gamma}$& $\varepsilon_{\gamma}^{l}$ \\
& $\mathrm{fm}$ & $\mathrm{fm}$ & $\mathrm{MeV}$ & $\mathrm{MeV}$ & $\mathrm{MeV}$ & $\mathrm{MeV}$ & $\mathrm{MeV}$ & $\mathrm{MeV}$ \\
\hline

$\frac{1}{2}$&0.5792 & 0.4232&-129.2468&-86.16453 &-49.7014&-34.6822&72.8400&57.4635 \\\hline
        1.0  & 1.1641& 0.4232&-129.2468&-86.16453 &-17.8702&-11.1380&44.9240&37.3018 \\\hline
\multicolumn{9}{|c|}{Nuclear matter properties at saturation density} \\
\hline
\multicolumn{1}{|c|}{$\gamma$}&\multicolumn{1}{|c|}{$\rho_0$ ($\mathrm{fm}^{-3}$)} & \multicolumn{2}{c|}{$e (\rho_0) $ (MeV)}
& \multicolumn{1}{c|}{$K (\rho_0)$ (MeV)} & \multicolumn{1}{c|}{$\frac{m^*}{m}(\rho_0,k_{f_0})$}
& \multicolumn{1}{c|}{$E_s (\rho_0)$ (MeV)} & \multicolumn{2}{c|}{$L (\rho_0)$ (MeV)} \\
\hline
\multicolumn{1}{|c|}{$\frac{1}{2}$}& \multicolumn{1}{|c|}{0.1610} & \multicolumn{2}{c|}{-16.0} & \multicolumn{1}{c|}{237.5}
& \multicolumn{1}{c|}{0.686} & \multicolumn{1}{c|}{33.0} & \multicolumn{2}{c|}{70.8} \\\hline
\multicolumn{1}{|c|}{1.0}& \multicolumn{1}{|c|}{0.1610} & \multicolumn{2}{c|}{-16.0} & \multicolumn{1}{c|}{282.2}
& \multicolumn{1}{c|}{0.686} & \multicolumn{1}{c|}{33.0} & \multicolumn{2}{c|}{72.8} \\\hline
\end{tabular}
\end{table}

\section{Results and Discussion} 
\label{Sec:res}

 In order to examine the convergence of the Taylor series expansion of the energy 
density, we calculate the $2^{nd}$ and  $4^{th}$ order contributions, given 
in equations ({\ref{eq12}}) and ({\ref{eq13}}), respectively, for our SEI in equation ({\ref{eq23}}) as a function of density.
The results 
are shown
in the Figure 2 for the EOS of stiffness $\gamma= 1/2$. 
The $4^{th}$ order contribution can 
be seen to be very small, at most represents 2\%  of the $2^{nd}$ order one over the
whole range of density, justifying the validity of the Taylor series expansion of the 
energy density. It may be mentioned here that only the kinetic and finite range exchange parts 
of the interactions contribute to the $4^{th}$ order term, and its functional dependence in the
case of SEI makes it negative beyond a density
$\rho $ $\approx$1.55 $fm^{-3}$. For the sake of comparison we have also plotted 
the symmetry energy density obtained under the empirical parabolic approximation (PA).
The results of the $2^{nd}$ order Taylor series expansion and the PA compare well differing 
at most within 2\% over the whole range of density. 
The equilibrium proton fraction has been calculated for the two EOSs of table 1 in the exact 
analytic case and in the $2^{nd}$ and $4^{th}$ order approximations of the Taylor series expansion 
of the energy density by solving the corresponding charge neutrality and $\beta-$equilibrium conditions. The
$\beta-$equilibrium conditions for the $2^{nd}$ and $4^{th}$ order approximations are given in equations 
(\ref {eq19}) and (\ref {eq20}), respectively. In the PA, the proton fraction is calculated 
from equation (\ref {eq19}) with $E_{sym,2}$ replaced by equation (\ref {eq15}).
The proton fraction for the EOS of $\gamma=$1/2 is shown as a function of density in the {\it left} 
panel of Figure 3. In the {\it right} panel, the asymmetric contribution to the nucleonic part of the energy density 
$S^{NSM}(\rho)=[H(\rho,Y_{p})-H(\rho)]$ in the $\beta-$equilibrated $n+p+e+\mu$ matter, for the four 
cases mentioned before, is shown as a function of density.
\begin{figure}
\vspace{0.6cm}
\begin{center}
\includegraphics[width=0.8\columnwidth,angle=-90]{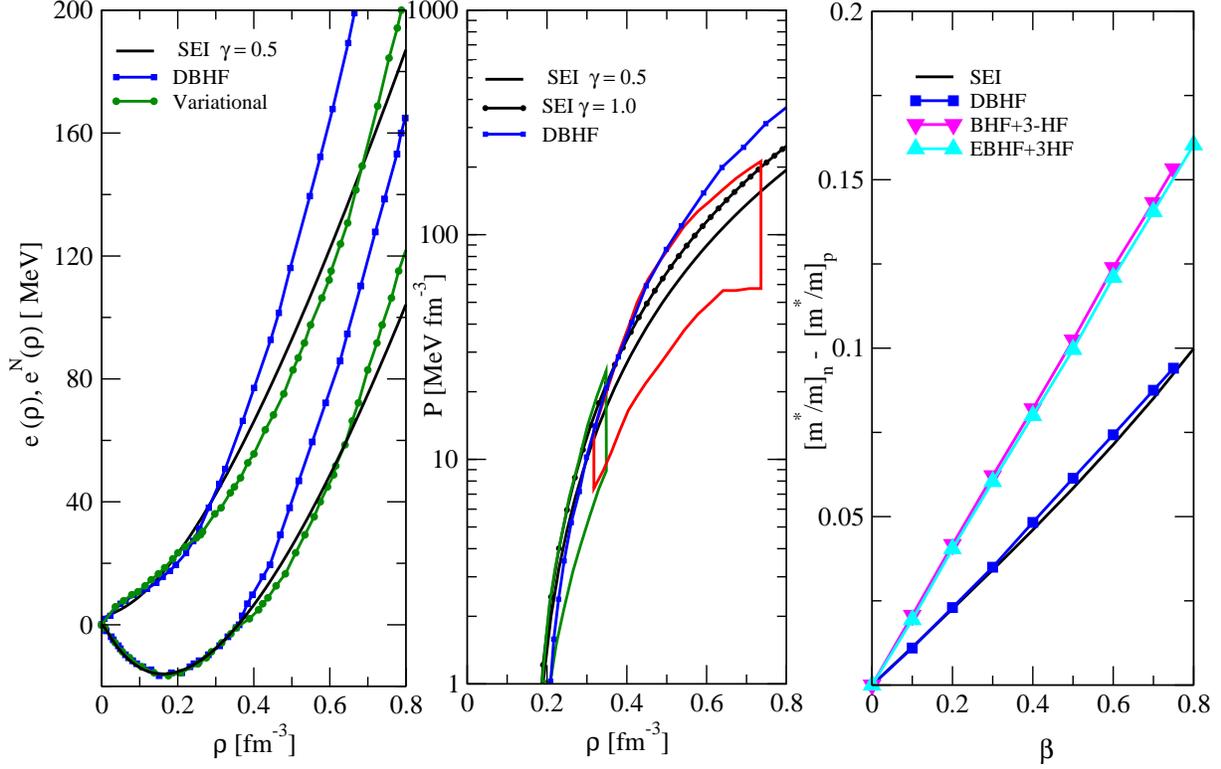}
\caption{{\it (color online)} {\it (left panel)} The energy per particle in SNM and PNM, $e(\rho)$ and $e^{N}(\rho)$, 
shown as a function of density $\rho$ for the EOS of SEI having $\gamma=1/2$. The corresponding results of 
microscopic DBHF \cite {sam10} and variational calculations using a realistic interaction \cite {akm98} have been given 
for comparison. {\it (middle panel)} Pressure-density relation in SNM of the two EOSs of SEI having $\gamma$=1/2 and 1 shown along with DBHF predictions \cite {sam10} and the allowed regions extracted from the heavy-ion collision studies (area within the red boundary) \cite{Danielz02} and the kaon production data (area within the green boundary) \cite{lynch09}. {\it (right panel)} The neutron and proton effective mass difference in normal nuclear matter is shown 
as a function of asymmetry $\beta$ for the SEI and compared with the predictions of the microscopic DBHF 
\cite {sam10}, BHF+3BF and EBHF+3BF \cite {zuo05} calculations.}
\label{Figure.0}
\end{center}
\end{figure}
\begin{figure}
\vspace{0.6cm}
\begin{center}
\includegraphics[width=0.8\columnwidth,angle=-90]{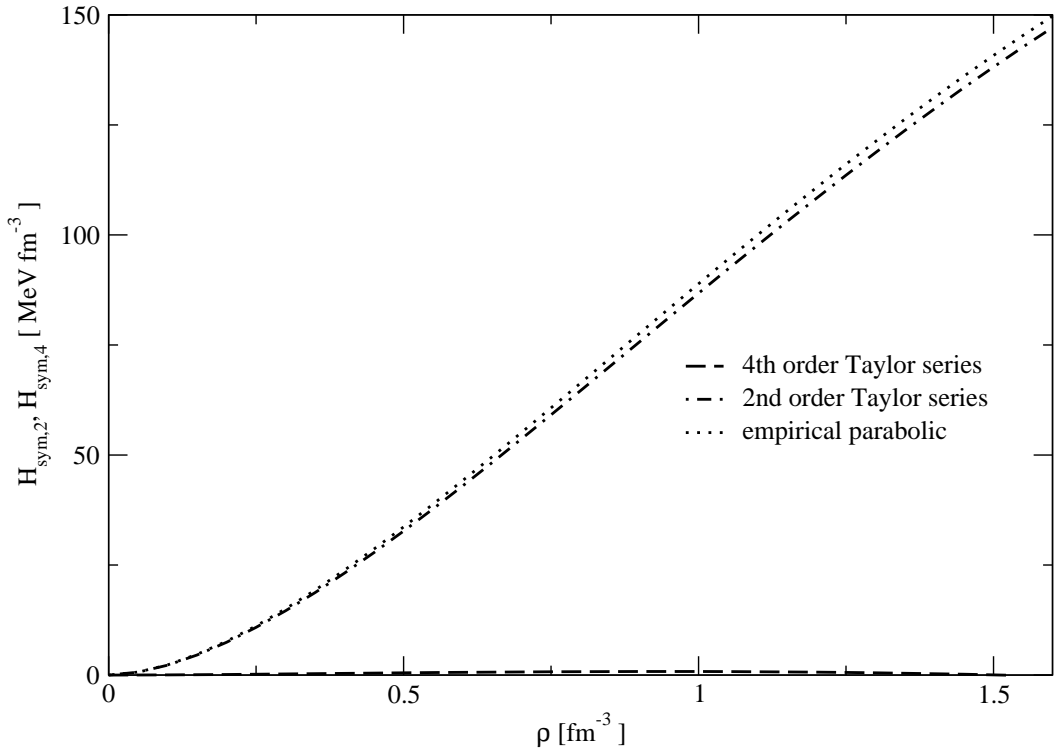}
\caption{ The $2^{nd}$ and $4^{th}$ order asymmetric contributions $H_{sym,2}$ and $H_{sym,4}$ 
to the energy density 
of ANM obtained under the Taylor expansion for the EOS of SEI corresponding to $\gamma$=1/2 is shown as a function 
of density $\rho$. The result of the asymmetric contribution obtained under the empirical parabolic approximation 
is also given for comparison. 
See text for details.}
\label{Figure.1}
\end{center}
\end{figure}
\begin{figure}
\vspace{0.6cm}
\begin{center}
\includegraphics[width=0.8\columnwidth,angle=-90]{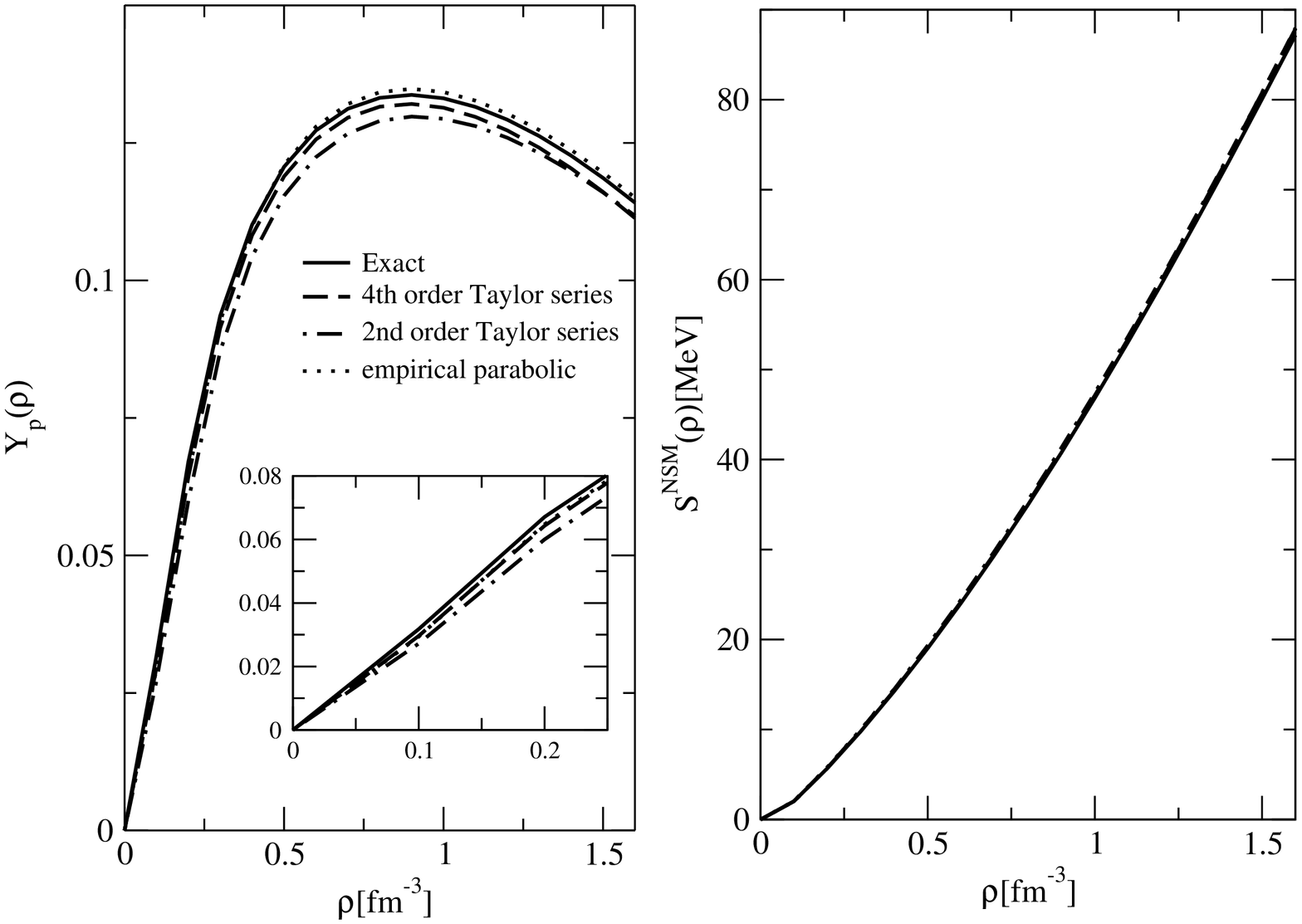}
\caption {{(\it {left}}) Proton fraction $Y_p(\rho)$ as a function of density in $\beta-$equilibrated $n+p+e+\mu$ matter (NSM) for the exact analytic expression of
the energy density, as well as for the $4^{th}$ and $2^{nd}$ order Taylor series approximations of the energy density and the empirical parabolic
approximation, for the EOS corresponding to $\gamma=$1/2 of table~1.
({\it {right}}) The contributions of the asymmetric part of the nucleonic energy densities, $S^{NSM}(\rho)$, in NSM
for the four approximations shown as a function of density. The labels of the curves are the same as in the {\it {left}} panel.}
\label{Figure.2}
\end{center}
\end{figure}
The curves of the proton fraction obtained under the different approximations lie very
close to each other over the whole density range, with the $2^{nd}$ order result
at the bottom. The small differences between the $2^{nd}$ order Taylor series
and the exact curves at different densities, is the measure of the cumulative contributions of all
higher order terms of the Taylor series to the equilibrium proton fraction
in $\beta-$stable matter. The curve for the empirical
parabolic approximation of equation (\ref {eq15}), where the symmetry energy is defined as the
difference between the energy per particle in PNM and SNM, however, remains above the three curves in
the density range beyond three times the normal nuclear matter density. The small
differences in the composition obtained under the various approximations to the exact energy density 
do not have any noticeable influence on the nucleonic part of the energy of the $\beta-$stable
matter. This is evident from the right panel of Figure 3, where the curves of the different approximations 
practically overlap with the curve of the exact result over the whole range of density.
The observations are similar for the EOS of table 1 corresponding to $\gamma=$1.

The mass-radius relationship in neutron stars is obtained 
by solving the TOV equations associated to the two EOS given in Table~1. For each EOS, 
the calculations are performed exactly and using the three approximations discussed in the text.
The corresponding results are displayed in Figure 4.
It can be seen that the $2^{nd}$ and $4^{th}$ order Taylor series approximations nicely
reproduce the exact result in the whole range of considered densities. This is also the case of  
the parabolic approach, although some small differences respect to the exact calculation appear 
for masses of the neutron stars below 1.5 $M_{\odot}$. The reason of the success of the different
approaches used in this work to reproduce the exact mass-radius relationship lies on the fact that using these
approximations the total (nucleonic+leptonic) energy density and pressure reproduce very accurately
the exact values, as it can be seen in the two panels of Figure~5. The largest differences at 
$\rho=$ 2 $\mathrm{fm}^{-3}$, where the exact energy density and pressure are about 3209 and 1781
$\mathrm{MeV} \mathrm{fm}^{-3}$, respectively, are less than 2 $\mathrm{MeV} \mathrm{fm}^{-3}$ for the 
energy density and 4 $\mathrm{MeV} \mathrm{fm}^{-3}$ for the pressure. 
In order to obtain the mass-radius relationship,
our EOS, defined in the core, has been supplemented from a density 0.0582 $\mathrm{fm}^{-3}$ down by
the EOS in the crust provided by the Baym-Bethe-Pethick calculation \cite {bayma,baymb}. However,
it should be pointed out that to join the EOS in the core computed with a given model with the EOS
of the crust calculated with a different model, may produce a sizeable effect on the radius of
the lightest neutron stars \cite {baldo14}, although this aspect does not change the conclusion of our present study.
\begin{figure}
\vspace{0.6cm}
\begin{center}
\includegraphics[width=0.8\columnwidth,angle=-90]{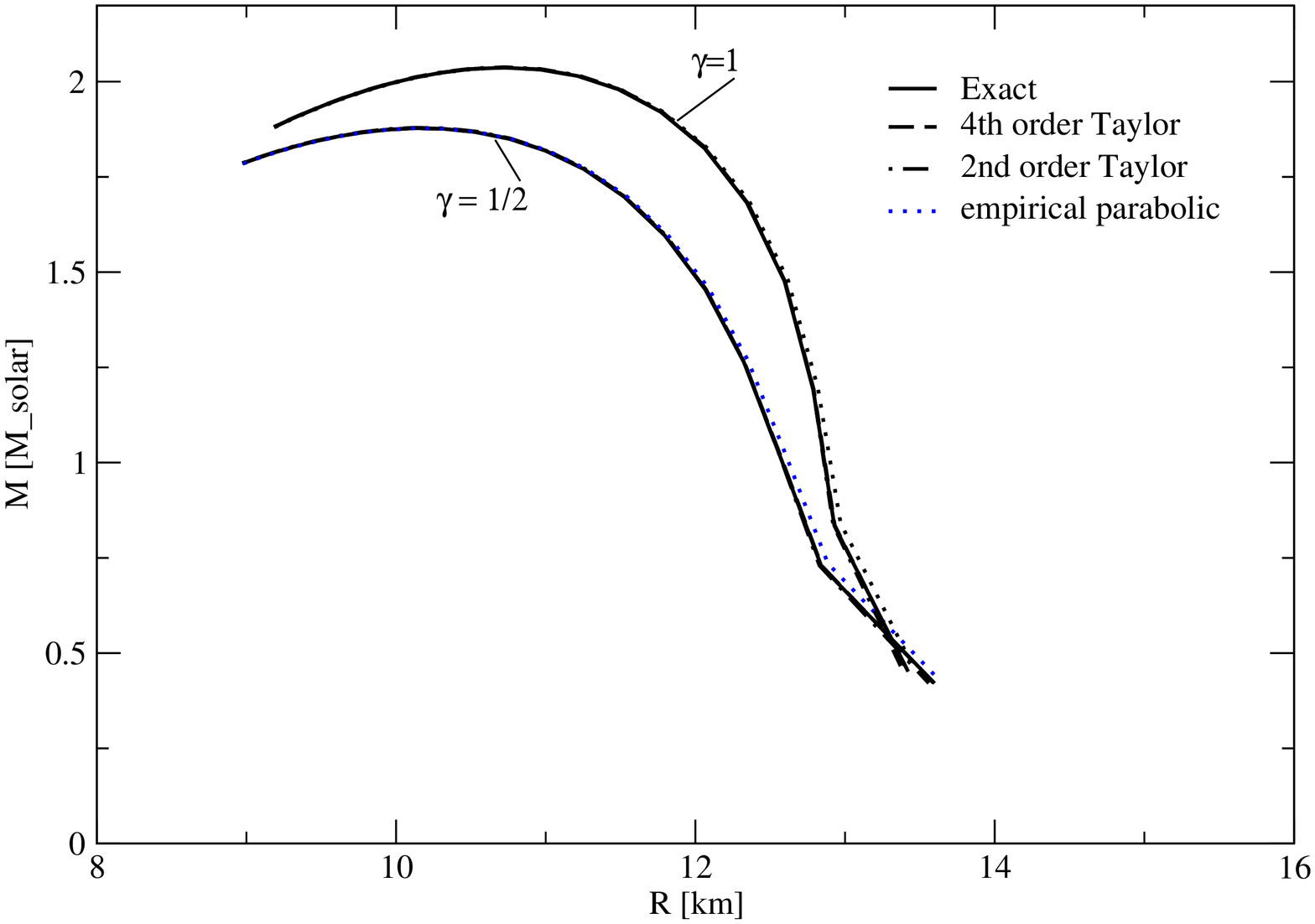}
\caption{({\it Color online}) Neutron star mass $M$ in units of the solar mass $M_{\odot}$ as a function of the radius $R$ in km 
obtained with the exact analytic expression of the 
energy density, the $4^{th}$ and $2^{nd}$ order Taylor expansion and empirical parabolic approximation of the energy density, for the two
sets of EOSs corresponding to $\gamma=$1/2 and $\gamma=$1 given in Table 1.}
\label{Figure.3}
\end{center}
\end{figure}%
\begin{figure}
\vspace{0.6cm}
\begin{center}
\includegraphics[width=0.8\columnwidth,angle=-90]{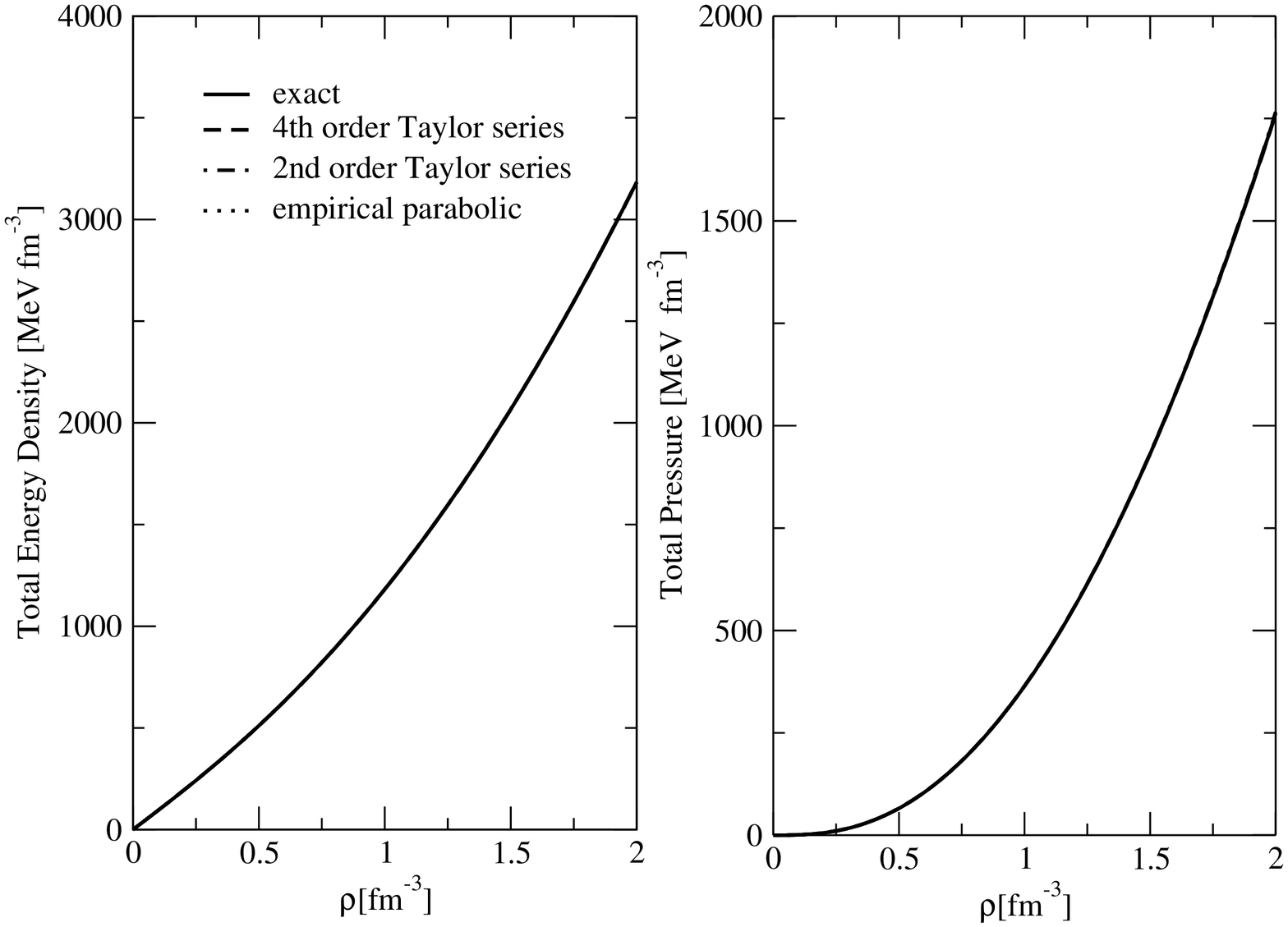}
\caption{({\it left}) Total energy density (nucleonic + leptonic), $H$, as a function of density 
for the exact analytic, the $4^{th}$, $2^{nd}$ order Taylor expansion and empirical parabolic approximations 
of the energy density corresponding to the EOS $\gamma=$1/2. 
({\it right} ) Same as the {\it left } panel but for the total pressure $P$. The labels of the curves are the same as in the {\it {left}} panel.}
\label{Figure.4}
\end{center}
\end{figure}
The maximum masses and radii of a NS obtained in the exact as well as in the approximated
calculations are 1.88 $M_{\odot}$ and 10.12 km and 2.04 $M_{\odot}$ and 10.73 km for the 
EOS of $\gamma=$1/2 and $\gamma=$1, respectively.
These predictions are compatible with the maximum mass 
constraint 1.97 $ \pm 0.04 M_{\odot}$, obtained from the estimation of mass
of the binary pulsar J1614-2230 by Demorest {\it et al.} \cite {dom10}.

We shall be using these two EOSs for 
the crust-core transition study in order to examine the influence of the
nuclear matter incompressibility on the transition density. 
The neutron star crust-core transition is another area where the contributions 
of higher order terms in the Taylor series expansion cannot be ignored, since the Taylor series 
approximation of the energy density of ANM can be misleading
in the prediction of important properties of NSs. One of such properties is the
crustal fraction of the moment of inertia, which critically depends on the pressure
at the transition density, as can be seen from equation (\ref {eq21}).

\begin{table}
\caption{Transition density $\rho_t$, pressure at transition density $P(\rho_t)$ and equilibrium proton fraction at transition density $Y_{p}(\rho_t)$ for the exact analytic case, $4^{th}$ and $2^{nd}$ order Taylor series approximations and the empirical PA of the two EOSs corresponding to $\gamma=$1/2 and  $\gamma=$1 of SEI given in table 1.
 The results of the calculations with two RMF lagrangians \cite{cai12} and two non-relativistic interactions (MDI and Skyrme R$_\sigma$) \cite{xu09} are given for comparison.}
\renewcommand{\tabcolsep}{0.18cm}
\renewcommand{\arraystretch}{1.2}
\begin{tabular}{|c|c|c|c|c|}
\hline

SEI            &         EOS										 & $\rho_{t} $   			 & P($\rho_{t})$   & $Y_{p}(\rho_{t})$  \\
               &                                 & $\mathrm{fm}^{-3}$   				 & $\mathrm{MeV} \mathrm{fm}^{-3}$ 	&        									\\\hline

$\gamma = 1/2$ &         Exact    							 &    0.0788           &  0.430        	&  0.0247               \\\hline
$\gamma = 1/2$ & $4^{th}$ order Taylor series    &    0.0919           & 	0.653					&  0.0271                       \\\hline
$\gamma = 1/2$ & $2^{nd}$ order Taylor series    &    0.0954           & 	0.715 				&  0.0258                   \\\hline
$\gamma = 1/2$ &   empirical parabolic           &    0.0953           & 	0.742					&  0.0279                   \\\hline
$\gamma = 1  $ &         Exact      						 &    0.0845           & 	0.468					&  0.0258               \\\hline
$\gamma = 1  $ & $4^{th}$ order Taylor series    &    0.0962           & 	0.682					&  0.0277             \\\hline
$\gamma = 1  $ & $2^{nd}$ order Taylor series    &    0.0994           & 	0.742					&  0.0263                  \\\hline
$\gamma = 1  $ &   empirical parabolic           &    0.0992           &	0.769					&  0.0284                      \\\hline \hline
FSUGold				 & $4^{th}$ order Taylor series    &    0.051            & 	0.321					&               \\
\cite{cai12}   & $2^{nd}$ order Taylor series    &    0.089            & 	1.316 				&                  \\\hline
IU-FSU				 & $4^{th}$ order Taylor series    &    0.077            & 	0.530					&            \\
\cite{cai12}    & $2^{nd}$ order Taylor series   &    0.090            & 	0.673 				&                     \\\hline
MDI$(x=0)$		 &         Exact    					  	 &    0.073            & 						&            \\
\cite{xu09}  	 & $2^{nd}$ order Taylor series    &    0.090           & 					&                     \\\hline
Skryme force $R_{\sigma}$ &  Exact                 &    0.066           & 	 0.316			&      0.0143               \\
                   &  $4^{th}$ order Taylor series &   0.089            &  0.766			&       0.0189              \\
 \cite{xu09}       & $2^{nd}$ order Taylor series&    0.093               & 	    0.898			&        0.0184             \\ \hline
\end{tabular}
\end{table}

The density at which the crust-core transition  takes place is calculated
from the respective stability conditions in equations (\ref {eq16}) and (\ref{eq17})
for the $2^{nd}$ and $4^{th}$ order Taylor series approximations. The 
corresponding result for the exact treatment of the energy density is also calculated using the 
stability condition expression in equation (\ref {eq8}).
 The pressure at the transition density $P(\rho_t)$ is a crucial quantity for the calculation of the crustal fraction of the moment of inertia used in the possible explanation of the observed glitches in pulsars (see equation (\ref{eq21})).
The results for the transition density
$\rho_t$, pressure at the transition density $P(\rho_t)$ and corresponding
proton fraction $Y_P(\rho_t)$ obtained from the three calculations in each 
of the two EOSs of table 1 are listed in table 2. For the sake of comparison, the results for the 
empirical PA for the two EOSs are given in the 
same table, along with other values calculated in the literature
using the FSUGold parameter set of the relativistic mean field model (RMF), and MDI and Skyrme interactions in non-relativistic
formulations  \cite {cai12,xu09}.  Though the results of the different parameter sets differ in magnitude, the same trend of predicting a
lower transition density (and lower pressure at the transition density) upon inclusion of the $4^{th}$ order term in comparison to the $2^{nd}$ order result
is found in all the sets of both the relativistic and non-relativistic calculations. Moreover, although the inclusion of the $4^{th}$ order term moves the transition density and pressure at the transition density in the right direction, the values are still far from the results obtained in the exact calculation. This clearly shows that in order to have the correct
prediction of the transition density of a given interaction, it would be necessary to
include all the terms, thereby implying the need for using the exact energy density expression. Comparison of the
results of $\rho_t$ in the $2^{nd}$ order Taylor series approximation and the
empirical PA shows that the predictions in both the cases are similar. 
The influence of the nuclear matter incompressibility can be seen 
from the results of the transition density for the two EOSs of SEI in table 2. The 
transition density, $\rho_t$, increases with an increase in the stiffness of 
nuclear matter. In SEI, an increase from 0.0788 $\mathrm{fm}^{-3}$ 
to 0.0845 $\mathrm{fm}^{-3}$ is obtained for the exact case as $K(\rho_0)$ increases from 
237 MeV ($\gamma=1/2$) to 282 MeV ($\gamma=1$). Similarly, the increase in the 
values of the transition density in the case of the $4^{th}$ order approximation is from 
0.0919 $\mathrm{fm}^{-3}$ to 0.0962 $\mathrm{fm}^{-3}$, and from 0.0954 $\mathrm{fm}^{-3}$ to 0.0994 $\mathrm{fm}^{-3}$ in the $2^{nd}$
order case, as $\gamma$ increases from $1/2$ to $1$. This is illustrated in Figure 6, where $V_{thermal}$ is plotted as a 
function of density in the close vicinity of the transition density for these three approximations, in the two EOSs 
having $\gamma=$1/2 and $\gamma=$1.

 The results of $\rho_{t}$ for the exact and the $2^{nd}$ and $4^{th}$ order Taylor expansion
of the energy density given in Table 2 suggest the possibility that the transition density obtained
using the Taylor series expansion might not have a convergent behaviour. In order to verify
this possibility we examine the behaviour of various properties, namely, energy per particle $e(\rho,Y_{p})$,
pressure $P^{N}(\rho,Y_{p})$ and $V_{thermal}$ computed  exactly and using the Taylor expansions and the
empirical PA of the energy density. For our analysis we have chosen two densities, $\rho = 0.08 $ fm$^{-3}$ and 1.2 fm$^{-3}$,
the former lies in the region of the crust-core transition while the latter corresponds to the NS central density
region. The  behaviour of $e(\rho,Y_{p})$ and $P^{N}(\rho,Y_{p})$ as a function of $Y_{p}$ is shown in Figures 7 and 8,
respectively. From these figures it can be seen that the exact results and the ones obtained with the different
approximations considered in this work are quite similar over a wide range of asymmetry in both the low and high density regions.
The small discrepancy between the results obtained exactly and with the Taylor expansion at very large asymmetry,
particularly in $P^{N}(\rho,Y_{p})$, may require some additional higher order terms in the Taylor expansion series
for having a perfect convergence with the exact results. However, a very different behaviour is found if the same study 
is performed with $V_{thermal}$, in particular in the crust-core transition density domain.
$V_{thermal}$ as a function of $Y_{p}$ is computed using equations({\ref {eq8}}),  
({\ref {eq16}}) and ({\ref {eq17}})
for the exact and $2^{nd}$ and $4^{th}$ order Taylor expanded cases, respectively. To obtain $V_{thermal}$ in the PA, 
we use equation({\ref {eq16}}) together with the definition for symmetry energy in equation ({\ref {eq15}}). The corresponding results, computed at our reference densities $\rho = 0.08$ fm$^{-3}$ and $\rho=1.2$ fm$^{-3}$, are displayed in the two panels of Figure 9. From the right
panel of figure 9 it can be seen that in the high density region the agreement 
between $V_{thermal}$ computed exactly and 
using the Taylor expansions and the PA of the energy density is similar to that found with $e(\rho,Y_{p})$ and $P^{N}(\rho,Y_{p})$. 
The agreement between $4^{th}$ order and exact results is very good for proton 
fractions $Y_{p}$ larger than 0.05. 
The small differences for low proton fractions with $Y_{p}$ smaller than 0.05 could be, in principle, accounted for
by considering additional higher order terms in Taylor series.
However, the situation dramatically changes in the low density regime with $\rho = 0.08$ fm$^{-3}$,
as it can be observed in the left panel of figure 9. For proton fractions $Y_{p}$ larger than 0.2 the agreement between the 
$V_{thermal}$ computed exactly and using the $4^{th}$ order Taylor expansion is good, but the exact and $4^{th}$ order 
predictions of $V_{thermal}$ exhibit a completely different behaviour when
the proton fraction decreases below 0.2. For these low values of the proton fraction the exact $V_{thermal}$ shows a 
stiff rising behaviour
for small values of $Y_{p}$.   
The curves of $V_{thermal}$ computed using the $2^{nd}$ and $4^{th}$ order approximations also show an increasing trend
when the proton fraction $Y_{p}$ decreases, however their slopes are clearly smaller than in the exact calculation,
which magnifies the differences between the exact and approximated calculations of $V_{thermal}$ for small proton
fractions. Although the curve of $V_{thermal}$ for the $4^{th}$ order approximation shows some improvements over its
$2^{nd}$ order counterpart, it cannot reproduce the stiff increasing shown by the exact $V_{thermal}$ curve. 
The results for the $V_{thermal}$ obtained under the empirical PA coincide
with the curves for the $2^{nd}$ order Taylor expansion of the energy density. Since the equilibrium proton fractions
$Y_{p}(\rho_{t})$ at the transition density for the different sets of SEI given in table 2 lie within the range 0.02-0.03,
where the behaviour of $V_{thermal}$ for the exact case is sharply increasing, it is unlikely that $V_{thermal}$ obtained
by including terms of higher order of the Taylor series expansion of the energy density could reproduce the results for
$\rho_{t}$ of the exact calculation in the case of our SEI. In order to understand the stiff rising behaviour of $V_{thermal}$ at high 
asymmetry in the low density region, the contributions of various parts of the EOS, namely, the kinetic, the finite range 
exchange ($\varepsilon_{ex}^{l}$ + $\varepsilon_{ex}^{ul}$) part and the zero range ($\varepsilon_{0}$ + $\varepsilon_{\gamma}$) part to 
$V_{thermal}$ have been calculated at the two reference densities, $\rho$ $=$ 0.08 fm$^{-3}$ and 1.2 fm$^{-3}$, as a function of $Y_p $. 
This is shown in the two panels of Figure 10, where the $Y_p$-dependence of the contributions of the kinetic and finite range exchange $(kin+exch)$ and of the
$\varepsilon_{0}$- and $\varepsilon_{\gamma}$- parts ($\varepsilon_{0}$+$\varepsilon_{\gamma}$) for the exact case and the $4^{th}$ order Taylor approximation calculated from equations (\ref{eq8}) and (\ref{eq17}), respectively, is displayed at these two reference densities. The contributions of the zero range $(\varepsilon_{0}$+$\varepsilon_{\gamma})$-part are the same for both the exact case and the $4^{th}$ order Taylor approximation.
But the $(kin+exch)$ contributions, which are negative in both cases, are different as $Y_p$ decreases, giving a sharp rising behaviour in the exact case. 
This behaviour of the $(kin+exch)$-part of the exact case at low values of $Y_p$ is a characteristic feature at all the densities. But the 
$(\varepsilon_{0}$+$\varepsilon_{\gamma})$-part that gives a high positive contribution at higher density, as can be seen from the right panel of figure 10, dominates and overshadows this rising feature of the $(kin+exch)$-part at high density. 
Therefore, the behaviour of $V_{thermal}$ for the exact case and $4^{th}$ order Taylor expansion is nearly similar at high density as has been obtained in the right panel of figure 10.
However, the $(kin+exch)$ and the $(\varepsilon_{0}$+$\varepsilon_{\gamma})$ contributions are opposite and comparable in magnitude at low values of the density 
as can be seen from the left panel of figure 10, thereby, the characteristic stiff rising behaviour of the $(kin+exch)$-part at low $Y_p$ values is manifested in $V_{thermal}$ for the exact case shown in the left panel of figure 9.

\begin{figure}
\vspace{0.6cm}
\begin{center}
\includegraphics[width=0.8\columnwidth,angle=-90]{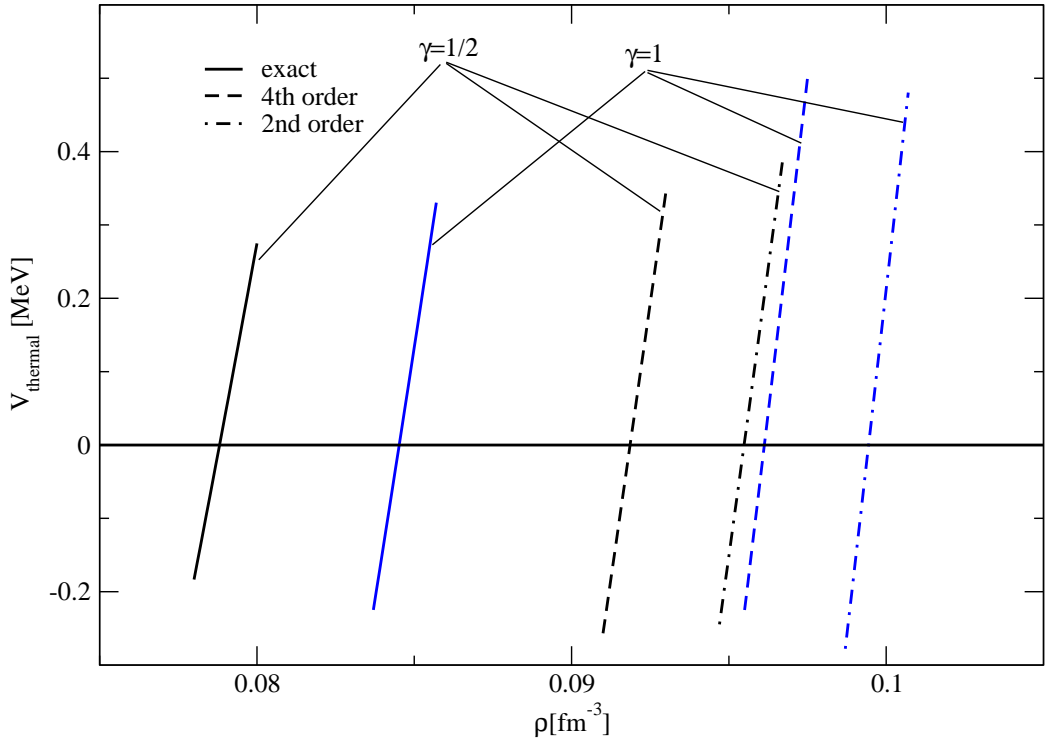}
\caption{({\it Color online}) $V_{thermal}$ in the region very close to the crust-core transition density
for the two sets of EOSs corresponding to $\gamma=$1/2 ($K(\rho_0)$=237 MeV) and
$\gamma=$1 ($K(\rho_0)$=282 MeV) given in Table 1. The results have been obtained
under the exact analytic and the $4^{th}$ and $2^{nd}$ order Taylor expansion of the energy density.}
\label{Figure.5}
\end{center}
\end{figure}
\begin{figure}
\vspace{0.6cm}
\begin{center}
\includegraphics[width=0.8\columnwidth,angle=-90]{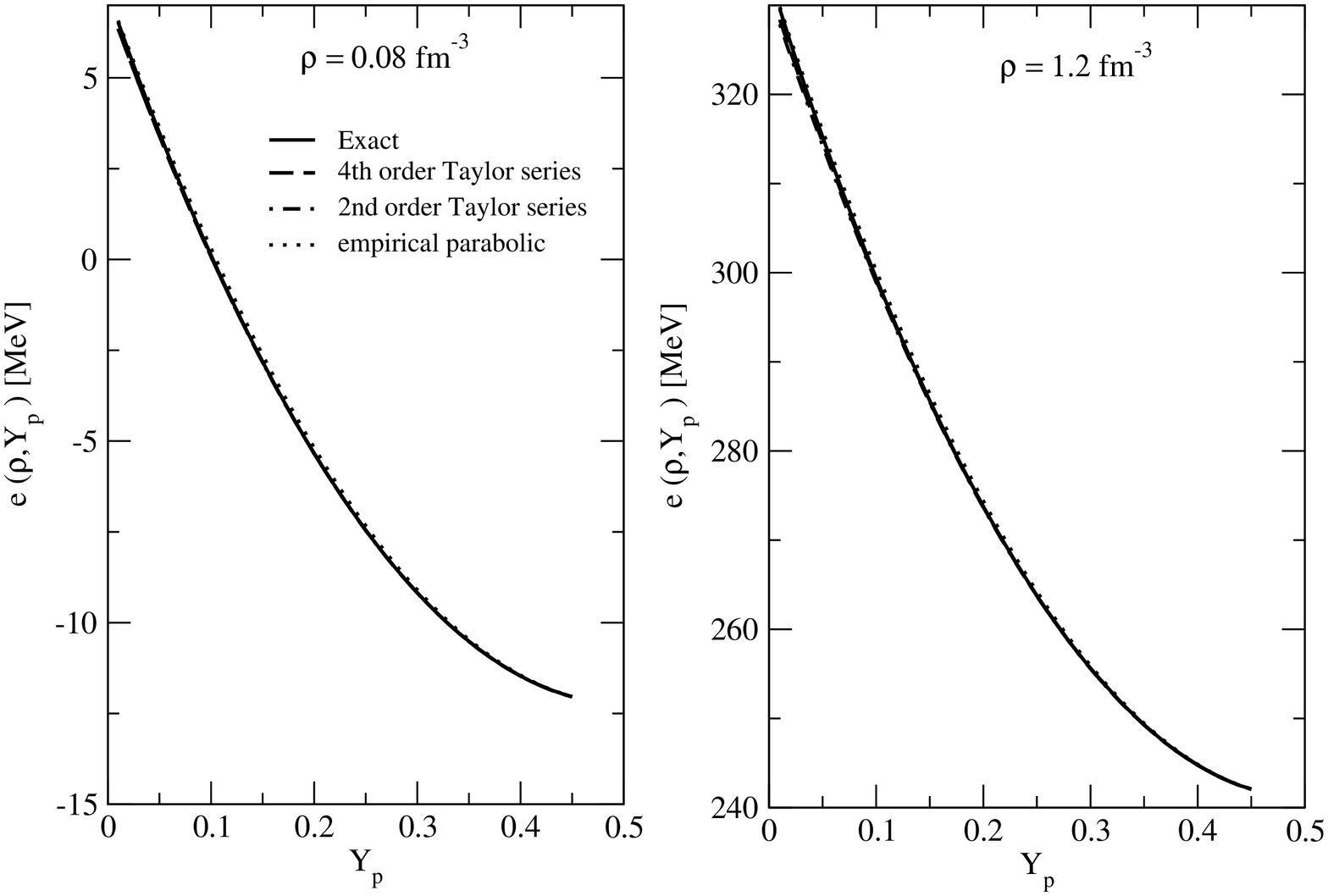}
\caption {{(\it {left})} The energy per particle $e(\rho,Y_{p})$ in ANM as a function of the proton fraction $Y_{p}$ at density $\rho=0.08$ $fm^{-3}$ for exact, $4^{th}$ and $2^{nd}$ order Taylor expansion and empirical PA of energy density.
{{(\it {right})} Same as the {{\it {left}} panel but at density $\rho=1.2$ $fm^{-3}$.}} The labels of the curves are the same as in the {\it {left}} panel.}
\label{Figure.6}
\end{center}
\end{figure}

The influence of the symmetry energy parameter $E_s(\rho_0)$ on $\rho_t$ is examined by 
calculating the transition density for two additional values $E_s(\rho_0)=30$ MeV and 36 MeV in the
EOS for $\gamma=$1/2, besides the value $E_s(\rho_0)=33$ MeV given in table 1.
The splitting of the two SNM parameters $\varepsilon_{0}$ and 
$\varepsilon_{\gamma}$ (recall equation (\ref{eq24})) in ANM for each value of $E_s(\rho_0)$, 
is made by searching for the value of $E_{s}^{'}(\rho_0)$ that satisfies the constraint of maximum asymmetric contribution of the 
nucleonic part in NSM, as discussed in subsection {\it {2.2}}. The values of $E_{s}^{'}(\rho_0)$ obtained for 
$E_s(\rho_0)$ =30 and 36 MeV are 21.64 MeV and 25.51 MeV, respectively. 
The splitting of $\varepsilon_{0}$ and $\varepsilon_{\gamma}$
in ANM does not have any influence on the predictions in SNM; 
however, it influences the stiffness of the symmetry energy curve.
The results for $\rho_t$ obtained in the exact case for $E_s(\rho_0)=$30 MeV and 36 MeV, are 0.0766 $\mathrm{fm}^{-3}$ and 0.0810 $\mathrm{fm}^{-3}$ respectively
(whereas the value of $\rho_t$ for $E_s(\rho_0)=$33 MeV is 0.0788 $\mathrm{fm}^{-3}$, cf. table 2).
Thus, an increase in the transition density with increasing $E_{s}(\rho_0)$ is predicted.
The same observation applies to the cases of the $4^{th}$ and $2^{nd}$ order Taylor series approximations.

The values of the slope parameter $L(\rho_0)$=3$E_s^{'}(\rho_0)$ for the two EOSs ($\gamma=$1/2 and $\gamma=$1), where $E_s^{'}(\rho_0)$ is determined from the constraint outlined in subsection {\it {2.2}}, are given in table 1.
The procedure of determination of the parameters in ANM adopted for SEI does not allow to make an independent study of the influence of
$E_s(\rho_0)$ and $E_s^{'}(\rho_0)$ (i.e. $L(\rho_0)$) on the transition density.
The influence of $L(\rho_0)$ on the transition density was examined in Refs.{\cite {xu09,mus12}}
for the MDI interaction (see figure 6 of Ref.\cite {xu09} and figure 4a and b of Ref.\cite {mus12}).
 In order to compare with the trend of MDI, in the present work we have
varied, in each of the two EOSs of table 1, the stiffness of the density dependence of the symmetry energy by starting from a low value of $E_s^{'}(\rho_0)$ and assigning increasing values to $E_s^{'}(\rho_0)$
(now, without imposing the constraint used in subsection {2.2} to fix $E_s^{'}(\rho_0)$ uniquely for a given value of $E_{s}(\rho_0)$).
The change of $\rho_t$ with the variation of
$E_s^{'}(\rho_0)$ is depicted in Figure 11. The trend obtained with SEI
for the $2^{nd}$ and $4^{th}$ order Taylor series expansion and the exact case is the same overall as obtained with MDI in Ref.\cite {xu09}.
It points out to the fact that the Taylor series approximated results cannot reproduce the trend of the exact calculation.
With the exact analytic expression of the energy density, $\rho_t$ shows a decreasing trend with an increase in the value of 
the slope parameter of the symmetry energy, whereas in the Taylor series approximations $\rho_t$
attains a minimum value and thereafter
follows a slow increasing trend with increasing slope parameter. In order to examine the influence of the 
functional form of the finite range part of the interaction on the transition density, calculations have been done with the Gaussian
form of $f(r)$ in equation (\ref{eq23})). Here we have constructed the EOS with the Gaussian form of $f(r)$ equivalent to the
$\gamma$=1/2 EOS of the Yukawa form given in table 1. For the same $\rho_0$ and $E_s(\rho_0)$ as of the Yukawa $\gamma$=1/2 EOS,
the $\gamma$ value required in the Gaussian form of SEI to predict the same nuclear matter incompressibility is $\gamma$=0.42.
Following the same parameter determination procedure, the parameters of this equivalent EOS were determined and the
crust-core transition density was calculated using the Gaussian EOS for the exact case and for the Taylor series $4^{th}$ and $2^{nd}$
order cases. The transition densities obtained in these cases are $\rho_t$=0.0788 $fm^{-3}$, 0.0916 $fm^{-3}$ and 0.0952 $fm^{-3}$,
respectively. These results are very similar to the ones obtained with the Yukawa form of the EOS with $\gamma$=1/2. The pressure and proton fraction at the transition density also are found to take similar values as in the Yukawa form of the EOS with $\gamma$=1/2. This shows that the use of Yukawa or Gaussian finite range form factors, at least in the SEI case, bears little influence on the prediction of the crust-core transition density in NSs.
\begin{table}
\caption{Neutron star radius $R$, mass $M$, crustal fraction of the moment of inertia $\frac{\Delta I}{I}$ and crustal thickness $\delta R$ as functions of the central density $\rho_c$ of the star. These quantities have been obtained for the EOS of $\gamma=$1/2 as defined in table 1. Results are given corresponding to the exact calculation as well as corresponding to the use of the $2^{nd}$ and $4^{th}$ order Taylor series expansion of the EOS.}
\renewcommand{\tabcolsep}{0.06cm}
\renewcommand{\arraystretch}{1.2}
\begin{tabular}{||c||c|c|c|c||c|c|c|c||c|c|c|c||}
\hline\hline
\multicolumn{1}{||c||}{}&\multicolumn{4}{|c||}{Exact}&\multicolumn{4}{|c||}{$4^{th}$ order Taylor series}&\multicolumn{4}{|c||}{$2^{nd}$ order Taylor series }\\\hline
$\rho_{c}$&$R$&	$M$ &$\frac{\Delta I}{I}$&$\delta R$&$R$&	$M$ &$\frac{\Delta I}{I}$&$\delta R$&$R$&	$M$ &$\frac{\Delta I}{I}$&$\delta R$\\\hline
$\mathrm{fm}^{-3}$&km&	$M_{\odot}$&fraction&km&km&	$M_{\odot}$&fraction&km&km&	$M_{\odot}$&fraction&km\\\hline
 2.00 &  8.986 &  1.785 & .0065  &.299  & 8.984 & 1.786 & .0089 & .326 & 8.982&    1.786&    .0096& 			     .332 \\\hline
 1.90 &  9.109 &  1.802 & .0067  &.306  & 9.106 & 1.802 & .0092 & .333 & 9.105&    1.803&    .0100&           .340\\\hline
 1.80 &  9.241&  1.818 & .0070  &.315  & 9.239 & 1.819 & .0097 & .342 & 9.238&    1.820&    .0105&           .350\\\hline
 1.70 &  9.386 &  1.834 & .0074  &.325  & 9.383 & 1.835 & .0102 & .353 & 9.382&    1.836&    .0110&           .361\\\hline
 1.60 &  9.543 &  1.849 & .0078  &.337  & 9.539 & 1.849 & .0107 & .366 & 9.539&    1.850&    .0116&           .374\\\hline
 1.50 &  9.713 &  1.861 & .0084  &.351  & 9.709 & 1.861 & .0115 & .382 & 9.709&    1.863&    .0124&           .390\\\hline
 1.40 & 9.898 &  1.870 & .0090  &.369  & 9.894 & 1.871 & .0124 & .401 & 9.895&    1.872&    .0134&           .410\\\hline
 1.30 & 10.100 &  1.876 & .0098  &.391  & 10.095& 1.876 & .0135 & .425 &10.096&    1.877&    .0145&           .434\\\hline
 1.20 & 10.319 &  1.875 & .0109  &.418  & 10.314& 1.875 & .0149 & .455 &10.314&    1.876&    .0160&           .464 \\\hline
 1.10 & 10.556 &  1.865 & .0122  &.453  & 10.550& 1.865 & .0167 & .493 &10.551&    1.866&    .0180&           .503 \\\hline
 1.00 & 10.812 &  1.843 & .0140  &.498  & 10.806& 1.843 & .0190 & .541 &10.807&    1.844&    .0205&           .553 \\\hline
  .90 & 11.087 &  1.804 & .0164  &.557  & 11.080& 1.803 & .0222 & .606 &11.081&    1.805&    .0240&           .618 \\\hline
  .80 & 11.379 &  1.741 & .0198  &.638  & 11.370& 1.740 & .0267 & .693 &11.371&    1.742&    .0288&           .707 \\\hline
  .70 & 11.682 &  1.646 & .0249  &.751  & 11.671& 1.645 & .0333 & .816 &11.672&    1.646&    .0357&           .832 \\\hline
  .60 & 11.987 &  1.507 & .0326  &.918  & 11.974& 1.505 & .0432 & .997 &11.974&    1.507&    .0462&          1.017 \\\hline
  .50 & 12.281 &  1.313 & .0452  &1.181 & 12.265& 1.310 & .0589 &1.281 &12.263&    1.312&    .0628&          1.306 \\\hline
  .40 & 12.554 &  1.053 & .0671  &1.632 & 12.532& 1.049 & .0855 &1.768 &12.527&    1.051&    .0904&          1.801 \\\hline
  .30 & 12.862 &   .732 & .1073  &2.531 & 12.833&  .729 & .1325 &2.734 &12.819&     .729&    .1386&          2.779 \\\hline
  .20 & 13.796 &   .392 & .1913  &4.919 & 13.762&  .389 & .2270 &5.282 &13.727&     .388&    .2355&          5.361 \\\hline \hline
\end{tabular}
\end{table}

The mass and radius of the neutron star as functions
of the central density $\rho_c$ of the star are calculated under the Taylor series approximations and the exact analytic case
in each of the two EOSs of table 1 by solving the corresponding TOV equations. The total
moment of inertia of the star is also calculated as a function of the central density from $I=J/{\Omega}$ for the considered cases.
Using the mass, radius, transition
density and pressure at the transition density of the NS,
the crustal fraction of the moment of inertia, $\frac {\Delta{I}}{I}$,
is calculated from equation (\ref {eq21}). Now using the calculated $I$, the contribution of the crust can be obtained. The results for the mass, radius, crustal fraction of the moment of inertia
and crustal thickness $\delta{R}$ (distance from the point of the transition density to the surface of the
neutron star) are given in Table 3 at different values of the NS central density $\rho_c$ for the EOS having $\gamma=$1/2.
Comparison of the results shows that although the use of the Taylor expansion of the energy density does not have any influence on the prediction of NS bulk properties such
as mass and radius,
its influence prominently manifests
in the predictions of the crustal fraction of the moment of inertia and the crustal thickness through the dependence on 
$\rho_t$ and $P(\rho_t)$. The two important 
quantities $\frac {\Delta{I}}{I}$ and $\delta{R}$ are clearly overestimated if one uses the Taylor expansion of the  energy density in the calculation
(see table 3). 
The comparison of the values of these two quantities computed exactly and with the $2^{nd}$ and $4^{th}$ order Taylor approximations and reported in table 3, shows a similar trend to the one exhibited by the transition density.

\begin{table}
\caption{The straight line equations for the radius $R$ of the Vela pulsar as a function of its mass $M$ in the $2^{nd}$ and $4^{th}$ order approximations and in the exact calculation for the two EOSs of table 1. Results of two additional EOSs having $\gamma=$1/2 but symmetry energy $E_{s}(\rho_0)=$36 and 30 MeV are also given.}
\renewcommand{\tabcolsep}{0.0991cm}
\renewcommand{\arraystretch}{1.2}
\begin{tabular}{|c|c|c|c|c|c|c|}\hline\hline
$\gamma$ & $E_{s}(\rho_{0})$ &  EOS				  &   $\rho_{t}$ &   $ P_{t}$  &     Vela Pulsar \\
		 			& 						     &  					  &   					    &   &   Radius constraint\\\hline
         & $\mathrm{MeV} $            &               & $\mathrm{fm}^{-3 }$  & MeV$\mathrm{fm}^{-3}$ &km\\\hline

1/2			 &   30&  Exact				   &0.0766&0.3235		&			   		 $R\geq 4.804+3.418 M/M_{\odot} $\\\hline
1/2			 &   33&  Exact					 &0.0788&0.4301		&					   $R\geq 4.386+3.335 M/M_{\odot} $\\
1/2			 &   33& $4^{th}$ order &0.0919 &0.629	  &					   $R\geq 3.884+3.270 M/M_{\odot} $\\
1/2			 &   33& $2^{nd}$ order &0.0954 &0.7145	&					   $R\geq 3.771+3.251 M/M_{\odot} $\\
1/2			 &   33& empirical PA    &0.0953&0.7428		&			 			 $R\geq 3.744+3.237 M/M_{\odot} $\\\hline
1/2			 &   36&  Exact				   &0.0810&0.5358		&			 			 $R\geq 4.105+3.257 M/M_{\odot} $\\\hline
1  			 &   33&  Exact				   &0.0845&0.4676		&			 			 $R\geq 4.252+3.340 M/M_{\odot} $\\
1  			 &   33& $4^{th}$ order	 &0.0962&0.6815		&			 			 $R\geq 3.951+3.298 M/M_{\odot} $\\
1  			 &   33& $2^{nd}$ order &0.0994&0.7416		&			  		 $R\geq 3.698+3.253 M/M_{\odot} $\\
1  			 &   33& empirical PA    &0.0992&0.7690		&			 		   $R\geq 3.673+3.239 M/M_{\odot} $\\\hline

\end{tabular}
\end{table}
\begin{figure}
\vspace{0.6cm}
\begin{center}
\includegraphics[width=0.8\columnwidth,angle=-90]{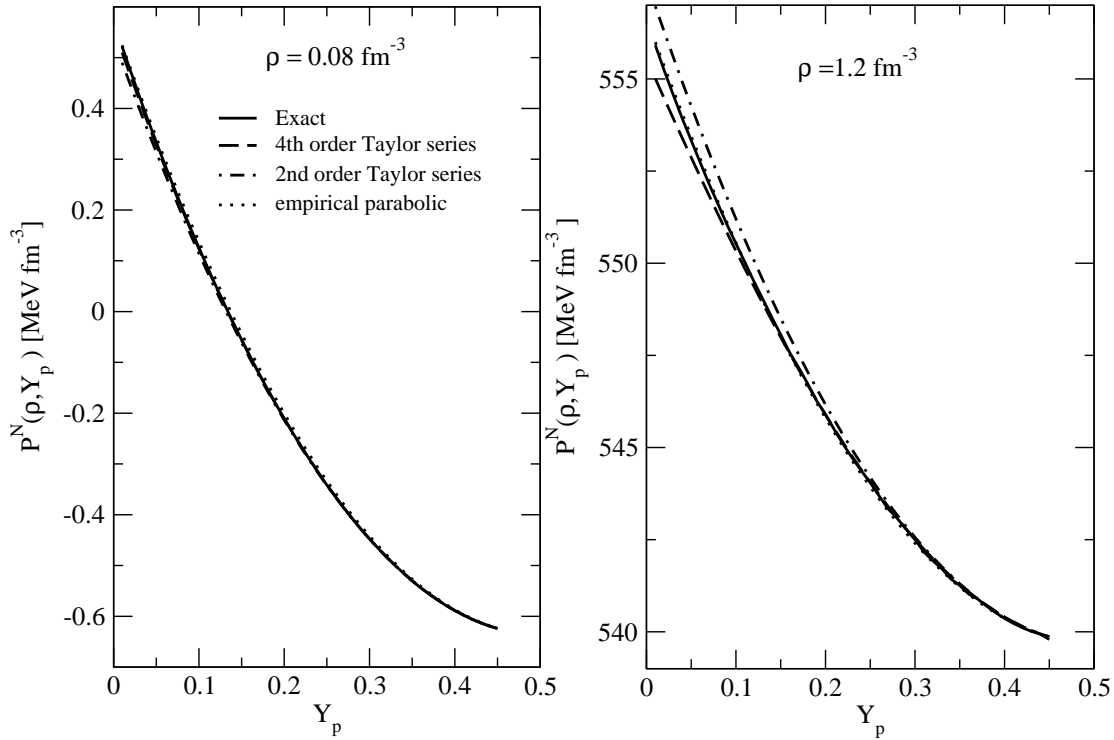}
\caption{ {(\it {left})} The nucleonic pressure $P^{N}(\rho,Y_{p})$ as a function of the proton fraction $Y_{p}$ at density
$\rho=0.08$ $fm^{-3}$ for exact, $4^{th}$ and $2^{nd}$ order Taylor expansion and empirical PA of energy density. {{(\it {right})} 
Same as the {{\it {left}} panel but at density $\rho=1.2$ $fm^{-3}$.}} The labels of the curves are the same as in the {\it {left}} panel.}
\label{Figure.7}
\end{center}
\end{figure}
\begin{figure}
\vspace{0.6cm}
\begin{center}
\includegraphics[width=0.8\columnwidth,angle=-90]{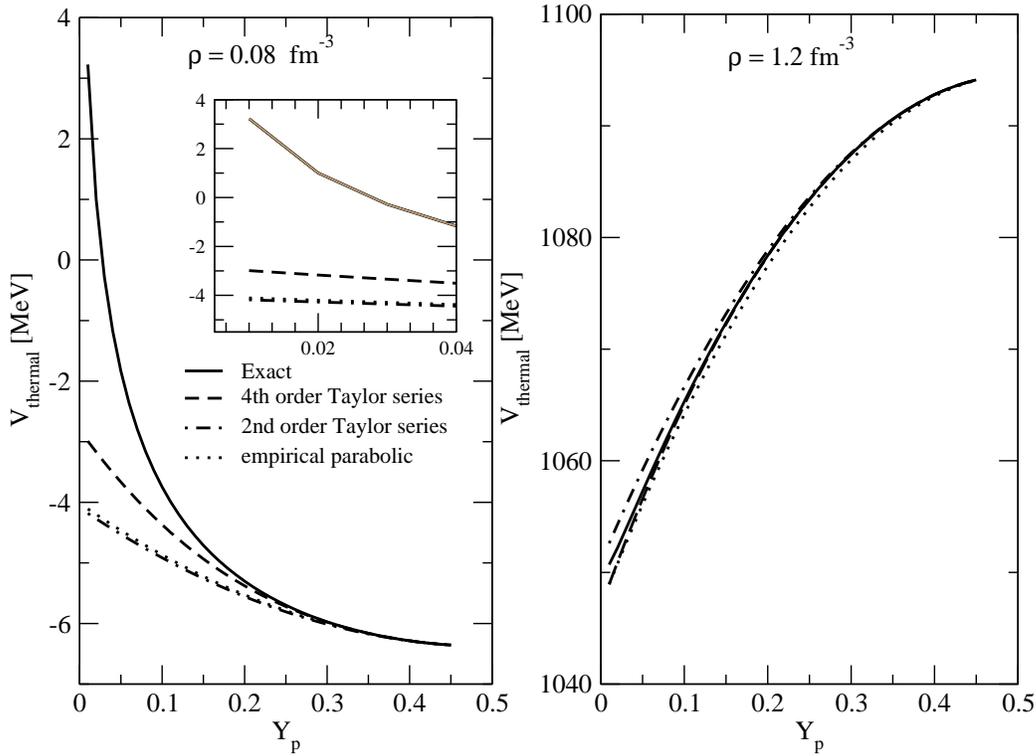}
\caption{{(\it {left})} $V_{thermal}$ as a function of the proton fraction $Y_{p}$ at density
$\rho=0.08$ $fm^{-3}$ for exact, $4^{th}$ and $2^{nd}$ order Taylor expansion and empirical PA of energy density. In the insert figure the same has been shown at the low values of proton fraction relevant to the crust-core transition region. For details see the text. 
{{(\it {right})} Same as the {{\it {left}} panel but at density $\rho=1.2$ $fm^{-3}$.}} The labels of the curves are the same as in the {\it {left}} panel.}
\label{Figure.8}
\end{center}
\end{figure}
\begin{figure}
\vspace{0.6cm}
\begin{center}
\includegraphics[width=0.8\columnwidth,angle=-90]{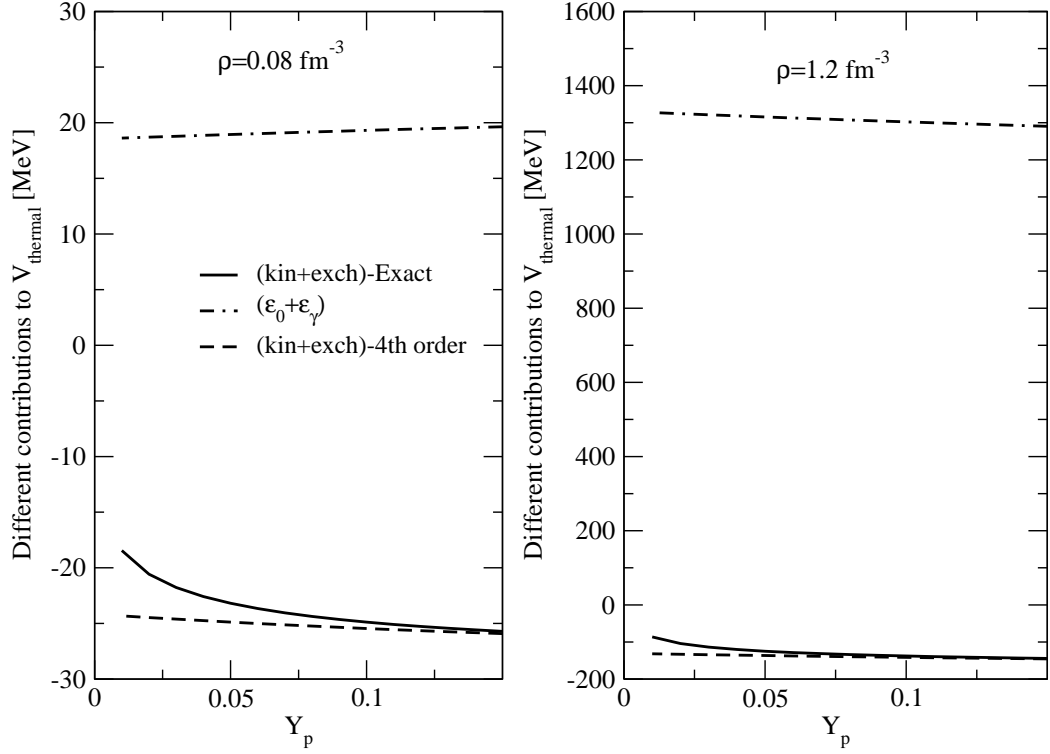}
\caption{{(\it {left})} Separate contributions from $(kin+exch)$- and $(\varepsilon_{0}$+$\varepsilon_{\gamma})$-parts to $V_{thermal}$ for exact and 
$4^{th}$ order Taylor expansion of the energy density as a function of proton fraction $Y_{p}$ at density
$\rho=0.08$ $fm^{-3}$. For details see the text. {{(\it {right})} Same as the {{\it {left}} panel but at density $\rho=1.2$ $fm^{-3}$.}} The labels of the curves are the same as in the {\it {left}} panel.}
\label{Figure.9}
\end{center}
\end{figure}
\begin{figure}
\vspace{0.6cm}
\begin{center}
\includegraphics[width=0.8\columnwidth,angle=-90]{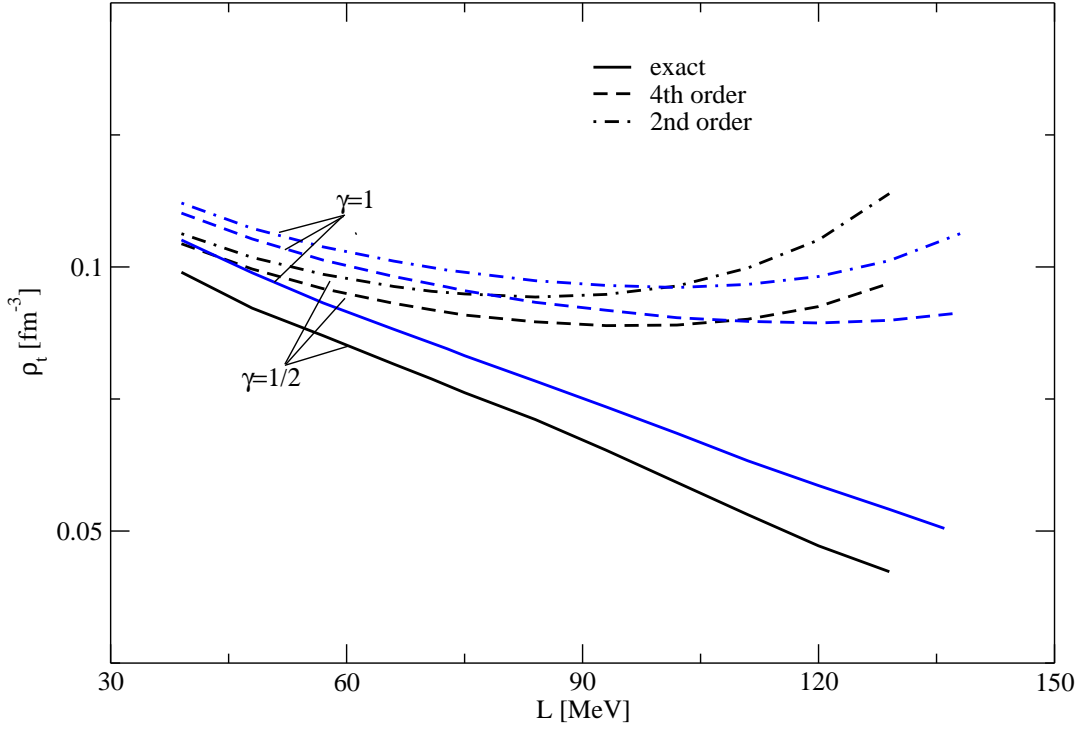}
\caption{({\it Color online}) Transition density $\rho_{t}$ for different values of the slope parameter $L(\rho_0)$ of the nuclear symmetry energy,
as obtained under the exact and the $4^{th}$ and $2^{nd}$ order Taylor expansion of the energy density for the two
EOSs corresponding to $\gamma$=1/2 and $\gamma$=1.}
\label{Figure.10}
\end{center}
\end{figure}

\begin{figure}
\vspace{0.6cm}
\begin{center}
\includegraphics[width=0.8\columnwidth,angle=-90]{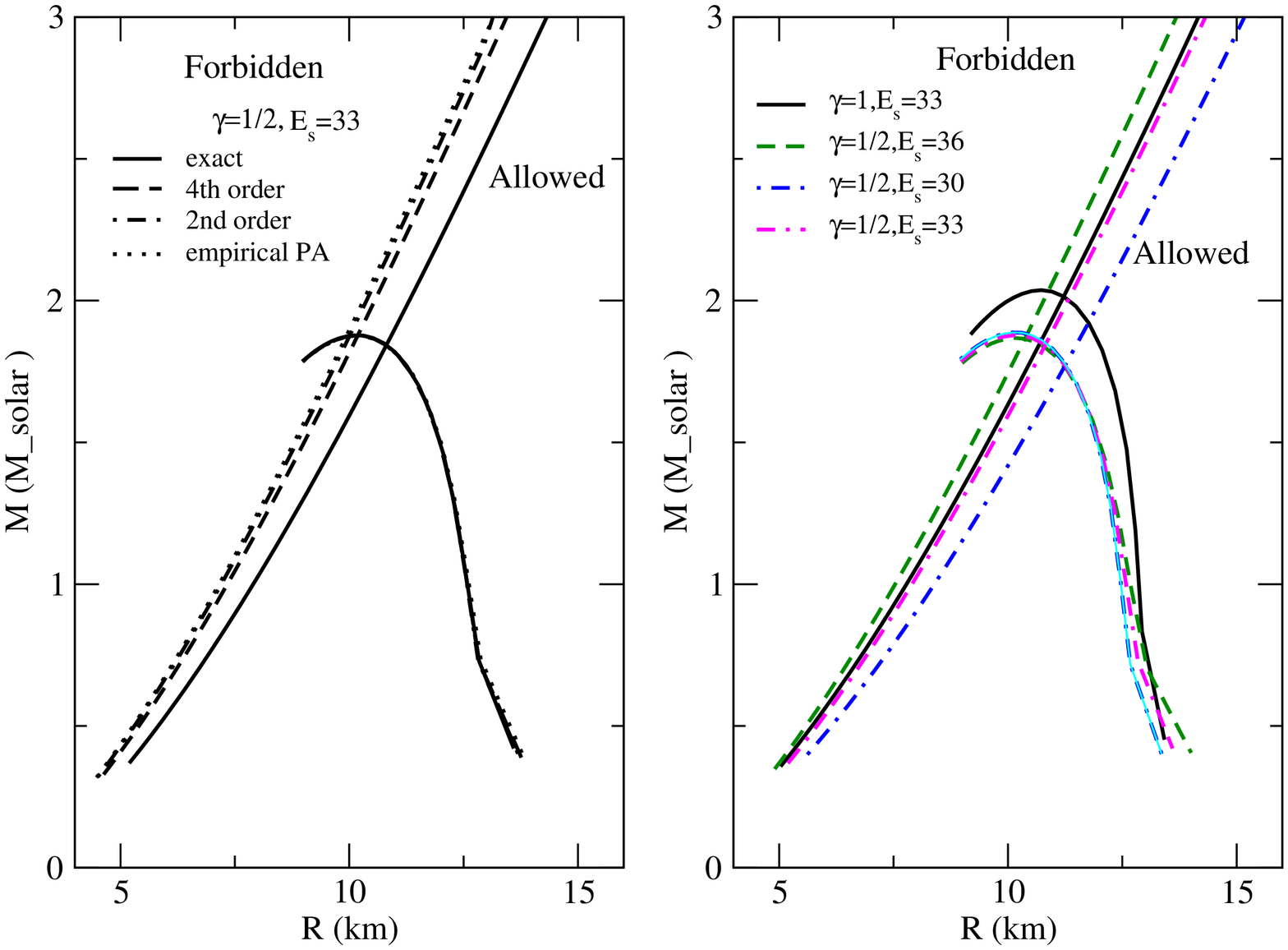}
\caption{({\it Color online}) ({\it left}) Radius $R$ as a function of the mass $M$ of the Vela pulsar for the EOS having $\gamma=$1/2 of table 1 under the various approximations along with their corresponding mass-radius predictions in NSs. The region {\it forbidden} corresponds to
the value of $\frac {\Delta{I}}{I}$ smaller than 0.014, the lower limit.
({\it right}) Same as in the {\it left} panel but for the exact treatment of the energy density for EOSs having $\gamma=$1/2 with symmetry energy $E_{s}(\rho_0)=$36, 33, 30 MeV, and $\gamma=$1 with symmetry energy $E_{s}(\rho_0)=$33 MeV.}
\label{Figure.11}
\end{center}
\end{figure}

We have  further examined the consequences of the Taylor series expansion of the energy density
on the prediction of the mass-radius constraint of the Vela pulsar. From the analysis of the
observed data on glitches of the Vela pulsar, the lower limit for 
$\frac {\Delta{I}}{I}$ is obtained to be 0.014 \cite {link99}. Using this condition in equation (\ref {eq21})
the radius of the Vela pulsar is expressed as a function of its mass, in the form of a straight line equation
for each considered EOS and approximation used. The results are given in Table 4.
The slope of the straight line relations for the $2^{nd}$ and $4^{th}$ order Taylor series approximations
decreases as compared to the exact calculation, predicting relatively smaller radius for the NS. This is shown in
the {\it left} panel of Figure 12, where the straight line curves depicting the predictions 
of the radius of the Vela pulsar as a function of the mass are plotted for the EOS having $\gamma=$1/2
and $E_{s}(\rho_0)=$33 MeV of table 1, under the different considered approximations. 
In the {\it right} panel of figure 12, similar results for the EOSs having different values of
$\gamma$ and $E_{s}(\rho_0)$ are given in the exact treatment of the energy density.
It can be seen from figure 12 that the predictions for the radius in the Vela pulsar for the
EOSs having $\gamma=$1/2 and $\gamma=$1 with the same $E_{s}(\rho_0)$ are very close to each other.
But the results show a marked sensitiveness on the value of the symmetry energy
$E_{s}(\rho_0)$. With increase (decrease) in $E_{s}(\rho_0)$ the predictions for the radius in the Vela pulsar decrease (increase), as can be seen in the {\it right} panel of figure~12 from the
results of the three EOSs having the same stiffness $\gamma=$1/2 but different symmetry energy parameter $E_{s}(\rho_0)=$36, 33 and 30 MeV.

\section{Summary and conclusions}
\label{Sec:con}
%
The influence of the higher order terms of the Taylor series expansion of the energy density
in the study of the crust-core transition in neutron stars is investigated. The calculations are
performed with the $2^{nd}$ and $4^{th}$ order Taylor expansions and with the exact treatment of 
the energy density for different equations of state based on finite range nuclear forces.
We find that the parabolic approximation of the energy in  ANM, often used in the nuclear calculations,
may be misleading as regards the predictions for the crust-core transition.
Even the inclusion of the $4^{th}$ order contribution 
cannot reproduce the results of the exact calculation in the case of SEI. This is due to the sharp rise of the slope of the $V_{thermal}$ computed exactly in the region of low density and low proton fraction that cannot be matched by the Taylor expanded calculations.
The transition density and pressure are
overestimated under the Taylor series approximation. Therefore, a Taylor expansion of the energy
in ANM shall lead to predictions for the properties of the neutron star sensitive to the physical conditions at 
the crust-core transition region that  do not correspond to the actual predictions of the EOS of the model. 
In this context, the crustal thickness and the
crustal fraction of the moment of inertia in neutron stars of different central density are found to take higher values
as compared to the exact result when the Taylor series approximation is used.

The analytic evaluation of further higher 
order terms in the Taylor expansion with a finite range interaction becomes a difficult task and prevents one to examine 
the convergence of the results to the exact prediction. Hence, where possible, the exact analytic 
expressions should be used in performing studies sensitive to the physical properties in the crust-core 
transition region. The stiffness parameter $\gamma$ in nuclear matter has the effect of increasing the transition density with an
increase in the incompressibility $K(\rho_0)$. Similarly, the transition density also increases with an increase in the value of the symmetry energy 
parameter $E_{s}(\rho_0)$. However, the stiffness of the symmetry energy has an opposite impact and the transition density takes up smaller values when
the value of the slope parameter $L$ is larger. All these effects have been examined in the case of the Vela pulsar, for which the lower 
limit of the crustal fraction of the moment of inertia has been ascertained from the study of the observed glitches. The nuclear matter incompressibility 
is found to have little influence in the prediction of the Vela pulsar radius. On the other hand, the symmetry energy parameter has 
a significant influence on the Vela pulsar radius which is predicted to take lower values for a larger symmetry energy parameter. 
Our present predictions for the crust-core transition density and pressure are based on the thermodynamical method.
A natural continuation of this work, to be done in the future, would be to extend the study reported in this paper using the more involved dynamical method.

\section{Appendix}
{\it {SEI with Yukawa form:}}
The energy density in asymmetric nuclear matter reads 
\begin{eqnarray}
H(\rho_n,\rho_p)&=&\frac{3\hslash^2}{10m}\left(k_n^2\rho_n+k_p^2\rho_p\right)
+\frac{\varepsilon_{0}^{l}}{2\rho_0}\left(\rho_n^2+\rho_p^2\right)
+\frac{\varepsilon_{0}^{ul}}{\rho_0}\rho_n\rho_p \nonumber \\
&&+\left[\frac{\varepsilon_{\gamma}^{l}}{2\rho_0^{\gamma+1}}\left(\rho_n^2+\rho_p^2\right)
+\frac{\varepsilon_{\gamma}^{ul}}{\rho_0^{\gamma+1}}\rho_n\rho_p\right]
\left(\frac{\rho({\bf R})}{1+b\rho({\bf R})}\right)^{\gamma} \nonumber \\
&&+\frac{\varepsilon_{ex}^{l}}{2\rho_0}(\rho_n^2 J(k_n) + \rho_p^2 J(k_p))\nonumber \\
&&+\frac{\varepsilon_{ex}^{ul}}{2\rho_0}\frac{1}{\pi^2}\left[\rho_n\int_0^{k_p}I(k,k_n)k^2dk
+\rho_p\int_0^{k_n}I(k,k_p)k^2dk\right] \nonumber \\
\end{eqnarray}
where the functions $J(k_i)$ and $I(k,k_i)$ with $k_i=(3 \pi^2 \rho_i)^{1/3}$
($i=n,p$) are given by
\begin{eqnarray}
J(k_i)=\bigg[\bigg(\frac{3}{32x_i^6}+\frac{9}{8x_i^4}\bigg)\ln(1+{4x_{i}^{2}})
-\frac {3}{8x_{i}^{4}}+\frac {9}{4x_{i}^{2}}-\frac {3}{x_{i}^{3}}\tan^{-1}(2x_{i}) \bigg], 
\label{eq10a}
\end{eqnarray}
and
\begin{eqnarray}
I(k,k_{i}) =&\frac{3(1+x^{2}_{i}-x^{2})}{8x^{3}_{i}x}\ln \bigg[\frac{1+(x+x_{i})^{2}}{1+(x-x_{i})^{2}} \bigg]\nonumber \\
&+\frac{3}{2x_{i}^{2}}-\frac{3}{2x_{i}^{2}}\bigg[  \tan^{-1}(x+x_{i})-\tan^{-1}(x-x_{i})\bigg],
\label{ppnma}
\end{eqnarray}
where $x_i$=$k_{i}/\Lambda$ (i=n,p), $x$=$k/\Lambda$  and $\Lambda=1/\alpha$.

 The expressions of the energy density for symmetric nuclear matter and pure neutron matter, $H(\rho)$ and $H^{N}(\rho)$, are
 \begin{eqnarray}
H(\rho) &=& \rho\, e(\rho)=\frac{3\hslash^2k_f^2\rho}{10m}
+\frac{(\varepsilon_{0}^{l}+\varepsilon_{0}^{ul})}{4\rho_0}\rho^2
+\frac{(\varepsilon_{\gamma}^{l}+\varepsilon_{\gamma}^{ul})}{4\rho_0^{\gamma+1}}
\rho^2\left(\frac{\rho({\bf R})}
{1+b\rho({\bf R})}\right)^{\gamma} \nonumber \\
 &&+\frac{(\varepsilon_{ex}^{l}+\varepsilon_{ex}^{ul})}{4\rho_0}
\rho^2\bigg[\bigg(\frac{3}{32x_f^6}+\frac{9}{8x_f^4}\bigg)\ln(1+{4x_{f}^{2}})
-\frac {3}{8x_{f}^{4}}+\frac {9}{4x_{f}^{2}}-\frac {3}{x_{f}^{3}}\tan^{-1}(2x_{f}) \bigg], \nonumber \\
\label{eq888}
\end{eqnarray}
\begin{eqnarray}
H^{N}(\rho) &=& \rho\, e^{N}(\rho)=\frac{3\hslash^2k_n^2\rho}{10m}
+\frac{\varepsilon_{0}^{l}}{4\rho_0}\rho^2
+\frac{\varepsilon_{\gamma}^{l}}{4\rho_0^{\gamma+1}}
\rho^2\left(\frac{\rho({\bf R})}
{1+b\rho({\bf R})}\right)^{\gamma} \nonumber \\
 &&+\frac{\varepsilon_{ex}^{l}}{4\rho_0}
\rho^2\bigg[\bigg(\frac{3}{32x_n^6}+\frac{9}{8x_n^4}\bigg)\ln(1+{4x_{n}^{2}})
-\frac {3}{8x_{n}^{4}}+\frac {9}{4x_{n}^{2}}-\frac {3}{x_{n}^{3}}\tan^{-1}(2x_{n}) \bigg], \nonumber \\
\label{eq101}
\end{eqnarray}
where $x_f$=$k_{f}/\Lambda$ .

The $2^{nd}$ and $4^{th}$ order parts of the Taylor series expansion of the energy density in asymmetric nuclear matter, defined in 
equations  (\ref {eq10}) and (\ref {eq11}), are given by
\begin{eqnarray}
 H_{sym,2}(\rho,Y_{p})&=\frac{ \hbar^{2}k^{2}_{f} \rho}{6m}+\frac{\rho^{2}}{4}\left(\frac{\varepsilon^{l}_{0}-\varepsilon^{ul}_{0}} {\rho_{0}}\right)
+\frac{\rho^{2}}{4}  \left( \frac{\varepsilon^{l}_{\gamma}-\varepsilon^{ul}_{\gamma}}{\rho_{0}^{\gamma + 1}}\right)\left(\frac{\rho}{1+b \rho}\right)^{\gamma}\nonumber\\
&+\frac{\rho^{2}}{4}  \left( \frac{\varepsilon^{l}_{ex}-\varepsilon^{ul}_{ex}}{\rho_{0}}\right)\bigg[\frac{\ln (1+4x^{2}_{f})}{4x^{2}_{f}}\bigg]\nonumber\\
&-\frac{\rho^{2}}{4}  \left( \frac{\varepsilon^{l}_{ex}+\varepsilon^{ul}_{ex}}{\rho_{0}}\right)\left[\left(\frac{1}{4x^{2}_{f}}+\frac{1}{8x^{4}_{f}}\right)\ln(1+4x^{2}_{f})-\frac{1}{2x^{2}_{f}}
\right]
\label{eq201}
 \end{eqnarray}
and
\begin{eqnarray}
 &H_{sym,4}(\rho,Y_{p})=\frac{ \hbar^{2}k^{2}_{f} \rho}{162m}\nonumber\\
 &+\frac{\rho^{2}}{108}  \left( \frac{\varepsilon^{l}_{ex}-\varepsilon^{ul}_{ex}}{\rho_{0}}\right) \left[\frac{\ln (1+4x^{2}_{f})}{x^{2}_{f}}-\frac{(8x^{4}_{f}+22x^{2}_{f}+4)}{(1+4x^{2}_{f})^{2}}\right]\nonumber\\
 &+\frac{\rho^{2}}{108}  \left( \frac{\varepsilon^{l}_{ex}+\varepsilon^{ul}_{ex}}{\rho_{0}}\right) \Bigg[
-\left(\frac{7}{8x^{4}_{f}}+ \frac {1}{x^{2}_{f}} \right)\ln(1+4x^{2}_{f})\nonumber\\
&+\frac{(16x^{6}_{f}+84x^{4}_{f}+50x^{2}_{f}+7)}{2x^{2}_{f}(1+4x^{2}_{f})^{2}} \Bigg]
 \label{eq202}
 \end{eqnarray} 
The proton and neutron chemical potentials, $\mu_p$ and $\mu_n$, and their derivatives used in equation (\ref {eq8}) take the expressions
\begin{eqnarray}
&\mu_{p}=\frac{\hbar^{2}k^{2}_{p}}{2 m}
+\left[\frac{\varepsilon^{l}_{0}}{\rho_{0}}+\frac{\varepsilon^{l}_{\gamma}}{\rho^{\gamma+1}_{0}}\left(\frac{\rho}{1+b \rho}\right)^{\gamma}\right ]\rho_{p}
+\left[\frac{\varepsilon^{ul}_{0}}{\rho_{0}}+\frac{\varepsilon^{ul}_{\gamma}}{\rho^{\gamma+1}_{0}}\left(\frac{\rho}{1+b \rho}\right)^{\gamma}\right ]\rho_{n}\nonumber\\
&+\varepsilon^{l}_{ex}\frac{\rho_{p}}{\rho_{0}}\left[ \frac{3\ln(1+4x^{2}_{p})}{8x^{4}_{p}}+\frac{3}{2x^{2}_{p}}-\frac{3 \tan^{-1}(2x_{p})}{2x^{3}_{p}} \right]\nonumber\\
&+\varepsilon^{ul}_{ex}\frac{\rho_{n}}{\rho_{0}} \Bigg[\frac{3(1+x^{2}_{n}-x^{2}_{p})}{8x_{p}x^{3}_{n}}
\ln\left[ \frac{1+(x_{p}+x_{n})^{2}}{1+(x_{p}-x_{n})^{2}}\right]
+\frac{3}{2x^{2}_{n}}\nonumber\\
&-\frac{3}{2x^{3}_{n}} (\tan^{-1} (x_{p}+x_{n})-\tan^{-1}(x_{p}-x_{n})) \Bigg]\nonumber\\
&+\left[\frac{\varepsilon^{l}_{\gamma}\left(\rho^{2}_{n}+\rho^{2}_{p}  \right)}{2\rho^{\gamma+1}_{0}}+ \frac{\varepsilon^{ul}_{\gamma}\left(\rho_{n}\rho_{p}  \right)}{\rho^{\gamma+1}_{0}}\right] \frac{\gamma \rho^{\gamma-1}}{(1+b\rho)^{\gamma+1}},
 \label{eq203}
 \end{eqnarray}
 \begin{eqnarray} 
&\mu_{n}=\frac{\hbar^{2}k^{2}_{n}}{2 m}
+\left[\frac{\varepsilon^{l}_{0}}{\rho_{0}}+\frac{\varepsilon^{l}_{\gamma}}{\rho^{\gamma+1}_{0}}\left(\frac{\rho}{1+b \rho}\right)^{\gamma}\right ]\rho_{n} +\left[\frac{\varepsilon^{ul}_{0}}{\rho_{0}}+\frac{\varepsilon^{ul}_{\gamma}}{\rho^{\gamma+1}_{0}}\left(\frac{\rho}{1+b \rho}\right)^{\gamma}\right ]\rho_{p}\nonumber\\
&+\varepsilon^{l}_{ex}\frac{\rho_{n}}{\rho_{0}}\left[ \frac{3\ln(1+4x^{2}_{n})}{8x^{4}_{n}}+\frac{3}{2x^{2}_{n}}-\frac{3 \tan^{-1}(2x_{n})}{2x^{3}_{n}} \right]\nonumber\\
&+\varepsilon^{ul}_{ex}\frac{\rho_{p}}{\rho_{0}} \Bigg[\frac{3(1+x^{2}_{p}-x^{2}_{n})}{8x_{n}x^{3}_{p}}
\ln\left[ \frac{1+(x_{p}+x_{n})^{2}}{1+(x_{p}-x_{n})^{2}}\right]
 +\frac{3}{2x^{2}_{p}}\nonumber\\
&-\frac{3}{2x^{3}_{p}} (\tan^{-1} (x_{n}+x_{p})-\tan^{-1}(x_{n}-x_{p})) \Bigg]\nonumber\\
&+\left[\frac{\varepsilon^{l}_{\gamma}\left(\rho^{2}_{n}+\rho^{2}_{p}  \right)}{2\rho^{\gamma+1}_{0}}+ \frac{\varepsilon^{ul}_{\gamma}\left(\rho_{n}\rho_{p}  \right)}{\rho^{\gamma+1}_{0}}\right] \frac{\gamma \rho^{\gamma-1}}{(1+b\rho)^{\gamma+1}},
 \label{eq204}
 \end{eqnarray}

  \begin{eqnarray} 
&\frac{\partial \mu_{p}}{\partial \rho_{p}}=\frac{\hbar^{2}k^{2}_{p}}{3 \rho_{p} m}
+\left[\frac{\varepsilon^{l}_{0}}{\rho_{0}}+\frac{\varepsilon^{l}_{\gamma}}{\rho^{\gamma+1}_{0}}\left(\frac{\rho}{1+b \rho}\right)^{\gamma}\right ]\nonumber\\
&+\left[\frac{\varepsilon^{l}_{\gamma}}{\rho^{\gamma+1}_{0}}\left(\frac{\rho_{p}}{\rho^{2}}\right)\gamma\left(\frac{\rho}{1+b \rho}\right)^{\gamma +1} \right]
+\left[\frac{\varepsilon^{ul}_{\gamma}}{\rho^{\gamma+1}_{0}}\left(\frac{\rho_{n}}{\rho^{2}}\right)\gamma\left(\frac{\rho}{1+b \rho}\right)^{\gamma +1} \right]\nonumber\\
&+\left(\frac{\varepsilon^{l}_{ex}}{\rho_{0}}\right)\left[\frac{-\ln(1+4x^{2}_{p})}{8x^{4}_{p}}+\frac{1}{2x^{2}_{p}}\right]\nonumber\\
&+\left(\frac{\varepsilon^{ul}_{ex}}{\rho_{0}}\right)\left(\frac{\rho_{n}}{\rho_{p}}\right)
\Bigg[\frac{-(1+x^{2}_{p}+x^{2}_{n})}{8x_{p}x^{3}_{n}}
\ln\left[ \frac{1+(x_{p}+x_{n})^{2}}{1+(x_{p}-x_{n})^{2}}\right]
\nonumber\\
&+\frac{(1+x^{2}_{n}-x^{2}_{p})^{2}+4x^{2}_{p}}{2x^{2}_{n} [1+(x_{p}+x_{n})^{2}][1+(x_{p}-x_{n})^{2}]}\Bigg]\nonumber\\
&+\left[\frac{\varepsilon^{l}_{\gamma}\rho_{p}+\varepsilon^{ul}_{\gamma}\rho_{n}}{\rho^{\gamma+1}_{0}} \right]\gamma \left( \frac{\rho^{\gamma-1}}{(1+b\rho)^{\gamma+1}}\right)\nonumber\\
&+\left[\frac{\varepsilon^{l}_{\gamma}\left(\rho^{2}_{n}+\rho^{2}_{p}  \right)}{2\rho^{\gamma+1}_{0}}+ \frac{\varepsilon^{ul}_{\gamma}\left(\rho_{n}\rho_{p}  \right)}{\rho^{\gamma+1}_{0}}
\right]\left[\frac{\gamma(\gamma-1)(1+b\rho)\rho^{\gamma-2}-b\gamma(\gamma+1)\rho^{\gamma-1}}{(1+b\rho)^{\gamma+2}} \right],
\nonumber \\
 \label{eq205}
 \end{eqnarray} 

   \begin{eqnarray} 
&\frac{\partial \mu_{n}}{\partial \rho_{n}}=\frac{\hbar^{2}k^{2}_{n}}{3 \rho_{n} m}
+\left[\frac{\varepsilon^{l}_{0}}{\rho_{0}}+\frac{\varepsilon^{l}_{\gamma}}{\rho^{\gamma+1}_{0}}\left(\frac{\rho}{1+b \rho}\right)^{\gamma}\right ]\nonumber\\
&+\left[\frac{\varepsilon^{l}_{\gamma}}{\rho^{\gamma+1}_{0}}\left(\frac{\rho_{n}}{\rho^{2}}\right)\gamma\left(\frac{\rho}{1+b \rho}\right)^{\gamma +1} \right]
+\left[\frac{\varepsilon^{ul}_{\gamma}}{\rho^{\gamma+1}_{0}}\left(\frac{\rho_{p}}{\rho^{2}}\right)\gamma\left(\frac{\rho}{1+b \rho}\right)^{\gamma +1} \right]\nonumber\\
&+\left(\frac{\varepsilon^{l}_{ex}}{\rho_{0}}\right)\left[\frac{-\ln(1+4x^{2}_{n}))}{8x^{4}_{n}}+\frac{1}{2x^{2}_{n}}\right]\nonumber\\
&+\left(\frac{\varepsilon^{ul}_{ex}}{\rho_{0}}\right)\left(\frac{\rho_{p}}{\rho_{n}}\right)
\Bigg[\frac{-(1+x^{2}_{n}+x^{2}_{p})}{8x_{n}x^{3}_{p}}
\ln\left[ \frac{1+(x_{p}+x_{n})^{2}}{1+(x_{p}-x_{n})^{2}}\right]
\nonumber\\
&+\frac{(1+x^{2}_{p}-x^{2}_{n})^{2}+4x^{2}_{n}}{2x^{2}_{p} [1+(x_{n}+x_{p})^{2}][1+(x_{n}-x_{p})^{2}]}\Bigg]\nonumber\\
&+\left[\frac{\varepsilon^{l}_{\gamma}\rho_{n}+\varepsilon^{ul}_{\gamma}\rho_{p}}{\rho^{\gamma+1}_{0}} \right]\gamma \left( \frac{\rho^{\gamma-1}}{(1+b\rho)^{\gamma+1}}\right)\nonumber\\
&+\left[\frac{\varepsilon^{l}_{\gamma}\left(\rho^{2}_{n}+\rho^{2}_{p}  \right)}{2\rho^{\gamma+1}_{0}}+ \frac{\varepsilon^{ul}_{\gamma}\left(\rho_{n}\rho_{p}  \right)}{\rho^{\gamma+1}_{0}}
\right]\left[\frac{\gamma(\gamma-1)(1+b\rho)\rho^{\gamma-2}-b\gamma(\gamma+1)\rho^{\gamma-1}}{(1+b\rho)^{\gamma+2}} \right],
\nonumber \\
\label{eq206}
 \end{eqnarray}

 \begin{eqnarray} 
&\frac{\partial \mu_{n}}{\partial \rho_{p}}=\frac{\partial \mu_{p}}{\partial \rho_{n}}=\frac{\varepsilon^{ul}_{0}}{\rho_{0}}+\left[\frac{(\varepsilon^{l}_{\gamma}+\varepsilon^{ul}_{\gamma})}{\rho^{\gamma+1}_{0}}\left(\frac{\gamma}{\rho}\right)\left(\frac{\rho}{1+b \rho}\right)^{\gamma +1} \right]\nonumber\\
&+\left[\frac{\varepsilon^{ul}_{\gamma}}{\rho^{\gamma+1}_{0}}\left(\frac{\rho}{1+b \rho}\right)^{\gamma} \right] \nonumber\\
&+\frac{\varepsilon^{ul}_{ex}}{\rho_{0}}   \left[  \frac{1}{4x_{p}x_{n}} \ln\left[\frac{1+(x_{p}+x_{n})^{2}}{1+(x_{p}-x_{n})^{2}}\right]\right] \nonumber\\
&+\left[\frac{\varepsilon^{l}_{\gamma}\left(\rho^{2}_{n}+\rho^{2}_{p}  \right)}{2\rho^{\gamma+1}_{0}}+ \frac{\varepsilon^{ul}_{\gamma}\left(\rho_{n}\rho_{p}  \right)}{\rho^{\gamma+1}_{0}}
\right]\left[\frac{\gamma(\gamma-1)(1+b\rho)\rho^{\gamma-2}-b\gamma(\gamma+1)\rho^{\gamma-1}}{(1+b\rho)^{\gamma+2}} \right].
\nonumber \\
\label{eq207}
 \end{eqnarray}

\section*{Acknowledgments}
The work of T. R. R is covered under the FIST program of School of Physics, Sambalpur University, India. 
The work of L. M. Robledo has been supported in part
by the Spanish MINECO Grants No. FPA2012-34694, and No. FIS2012-34479
and by the Consolider-Ingenio 2010 Program MULTIDARK CSD2009-00064.
M. C. and X. V. acknowledge partial support from Grant No.\ FIS2014-54672-P from the 
Spanish MINECO and FEDER, Grant No.\ 2014SGR-401 from Generalitat de Catalunya, 
the Consolider-Ingenio 2010 Programme CPAN CSD2007-00042, and the project 
MDM-2014-0369 of ICCUB (Unidad de Excelencia Mar\'{\i}a de Maeztu) from MINECO.
The ``NewCompStar'' COST Action MP1304 is also acknowledged. We also thank the
referees for their valuable suggestions to improve the paper to the present form.

\end{document}